\documentclass[fleqn,usenatbib]{mnras}

\usepackage{newtxtext,newtxmath}

\usepackage[T1]{fontenc}

\usepackage{multirow}

\DeclareRobustCommand{\VAN}[3]{#2}
\let\VANthebibliography\thebibliography
\def\thebibliography{\DeclareRobustCommand{\VAN}[3]{##3}\VANthebibliography}

\usepackage{graphicx}	
\usepackage{amsmath}	

\usepackage{booktabs}
\usepackage{subcaption}
\usepackage{xspace}
\usepackage{xcolor}

\newcommand{\kms}{\mathrel{\mathrm{km/s}}}
\newcommand{\msun}{\mathrel{\mathrm{M}_\odot}}

\newcommand{\fmix}{\mathrel{\mathrm{f}_{\rm mix}}}
\newcommand{\mlens}{\mathrel{\mathrm{M}_{\rm lens}}}
\newcommand{\vcm}{\mathrel{\mathrm{v}_\textrm{cm}}}
\newcommand{\taueff}{\mathrel{\tau_\textrm{eff}}}
\newcommand{\ncov}{\mathrel{\mathrm{n}_{\rm cov}}}
\newcommand{\nrec}{\mathrel{\mathrm{n}_{\rm rec}}}
\newcommand{\ntot}{\mathrel{\mathrm{n}_{\rm tot}}}
\newcommand{\msource}{\mathrel{\mathrm{m}_\mathrm{source}}}
\newcommand{\msourcesl}{\mathrel{\mathrm{m}_\mathrm{source,SL}}}
\newcommand{\musl}{\mathrel{\mu_\mathrm{sl}}}
\newcommand{\muslmax}{\mathrel{\mu_\mathrm{sl,max}}}
\newcommand{\porb}{\mathrel{\mathrm{P}_\mathrm{orb}}}

\newcommand{\ms}{\mathrel{\mathrm{M}_\mathrm{source}}}
\newcommand{\ml}{\mathrel{\mathrm{M}_\mathrm{lens}}}

\newcommand{\mlim}{\mathrel{\mathrm{m}_\mathrm{lim}}}
\newcommand{\sigmanoise}{\mathrel{\sigma_\mathrm{noise}}}

\newcommand{\gbhs}{\ensuremath{\mathcal{G}_\mathrm{BH,short}}\xspace}
\newcommand{\gbhl}{\ensuremath{\mathcal{G}_\mathrm{BH,long}}\xspace}
\newcommand{\gns}{\ensuremath{\mathcal{G}_\mathrm{NS}}\xspace}

\newcommand{\ensl}{\ensuremath{\mathbb{E}[\mathcal{N}_\mathrm{SL}]}}

\newcommand{\startrack}{\texttt{startrack}\xspace}
\newcommand{\cosmic}{\texttt{COSMIC}\xspace}

\title[Self-lensing binaries as probes of Supernova physics]{Self-lensing binaries as probes of Supernova physics}

\author[Wiktorowicz et al.]{Grzegorz Wiktorowicz$^{1}$\thanks{E-mail: gwiktoro@camk.edu.pl}, 
Matthew Middleton$^{2}$, Aleksandra Olejak$^3$, Cordelia Dashwood-Brown$^{2}$, 
\newauthor Madeleine-Mai Ward$^2$, Adam Ingram$^2$\\  
\\  
     $^1$ Nicolaus Copernicus Astronomical Center, Polish Academy of Sciences, Bartycka 18, 00-716 Warsaw, Poland\\
     $^2$ School of Physics \& Astronomy, University of Southampton, Southampton, Southampton SO17 1BJ, UK\\
     $^3$ Max Planck Institute for Astrophysics, Karl-Schwarzschild-Straße 1, 85748 Garching b. München, Germany\\
     $^4$ School of Mathematics, Statistics, and Physics, Newcastle University, Newcastle upon Tyne, NE1 7RU, UK \\
}

\date{Accepted XXX. Received YYY; in original form ZZZ}

\pubyear{2025}

\begin{document}
\label{firstpage}
\pagerange{\pageref{firstpage}--\pageref{lastpage}}
\maketitle

\begin{abstract}

Self-lensing (SL) in binary systems has the potential to provide a unique observational window into the Galactic population of compact objects. Using the \startrack and \cosmic population synthesis codes, we investigate how different supernova mechanisms affect the observable population of SL systems, with particular attention to the mass gap (2–-$5\msun$) in compact object distributions. We test three supernova remnant formation models with different convective growth timescales ($\fmix$ = 0.5, 1.0, and 4.0), simulating SL binary systems across the Galactic disk and bulge. We identify distinct groupings of SL sources based on lens mass and Einstein crossing time, clearly differentiating neutron star from black hole systems and close from wide orbits. Notably, the delayed $\fmix = 0.5$ model predicts a significantly higher fraction of systems with lens masses in the mass gap region (up to $\sim10$ times more for certain surveys), suggesting that SL observations could help constrain this controversial population. Our analysis reveals a strong preference for systems with low centre-of-mass velocities ($\vcm\leq20\kms$) across all models, resulting primarily from physical processes governing compact object formation and binary survival. While many potential detections will have limited observational coverage, ZTF is predicted to yield several dozen well-covered systems that should enable detailed characterization. When applying simple detection criteria including photometric precision and signal-to-noise requirements, predicted rates decrease by approximately two orders of magnitude, but still yield up to a few tens of expected detections for LSST and ZTF in the Galactic disk population. 

\end{abstract}


\section{Introduction}

The vast majority of the Galactic population of binary compact objects are expected to be part of relatively wide, non-interacting binary systems \citep{Wiktorowicz2019,Olejak2020,VignaGomez2023b}. Some fraction of these are systems with non-compact companions, typically low-mass, main sequence stars due to their long lifetime. Self-lensing (SL) occurs when a compact lens in a binary system transits its normal (optically bright) companion star. The result is an increase in light, the shape and amplitude of which depends on the mass of the lens, binary separation, impact parameter, projected size (radius) of the companion star and limb darkening \citep{Agol2003}. SL has been used to locate five white dwarf lenses to date, all within Kepler data \citep{Kruse1404,Kawahara1803,Masuda2019} and it has been suggested that quiescent (non-accreting) neutron stars (NSs) and black holes (BHs) may also be detectable given the high cadence of new surveys, especially within the all-sky  surveys of ZTF and the Vera C Rubin LSST (\citealt{Wiktorowicz_2021}, see also \citealt{Yamaguchi2024}). Recent detailed simulations of TESS observations have demonstrated detection efficiencies of $4$--$7$\% for SL systems with compact companions from TESS Candidate Target List stars \citep{Sedighe2412}. 

The potential of SL is only now being realized. Not only does the method not suffer from the biases inherent in other techniques which would tend to yield detections of high mass compact objects (thus precluding access to the low-mass gap), but the numbers are potentially much larger than known XRBs, permitting sensitive tests of supernova (SN) physics via correlations with the kick velocity \citep{Gandhi1905}. In addition, as most lenses found this way are pristine, having never accreted matter from a companion star \citep{Wiktorowicz_2021}, accessing their natal properties (mass and spin) can provide unique observational constraints on their formation mechanism.

Recent years have seen the first predictions for the population of SL systems in the Milky Way emerge. These have relied on various iterations of binary population synthesis codes from \cite{Masuda2019} and later \citet[hereafter W21]{Wiktorowicz_2021}. The latter used the \startrack\ population synthesis code \citep{Belczynski2008a}, a realistic model for the Milky Way, and included NS lenses for the first time. In W21, the simulated population was folded through the key instrumental properties of the surveys, assuming a range of conditions, namely the Initial Mass Function (IMF), star formation history and assuming the `Rapid' SN model \citep{Fryer1204,Belczynski1209}. One of the main features of this SN model is that it reproduces the observational mass gap between NSs and BHs \citep[around $2$--$5\msun$;][]{Bailyn9805}. However, observations -- both from gravitational waves \citep{Abbott2006} and spectroscopy \citep{Jayasinghe2106} -- have called into question the presence of such a gap. Should the gap truly contain systems, this may point towards a `Delayed' model for SN, where the growth timescale of the Rayleigh-Taylor/convective instability which launches the explosion is relatively long \citep{Belczynski1209}. In this paper, we build on the results obtained in W21 to explore the impact of new developments in SN engine physics \citep{Fryer2022} and explore the main characteristics of the SL population, including the kick velocity distributions and -- most importantly -- whether we can recover objects in the mass gap which would point towards a specific SN mechanism being dominant.

We follow the procedure outlined in W21 to calculate the expected population of SL objects in our synthetic populations of binaries. We choose to focus our analysis on those surveys which have a high potential for detecting SL events due to their sky coverage, sensitivity and cadence. Specifically, we include only those same survey instruments considered in W21: TESS, ZTF, and LSST (we refer the reader to W21 for a discussion of the various assumed instrumental characteristics such as duration, assumed cadence, saturation etc).

\section{Methods}

For this research, we utilized both the \startrack and \cosmic population synthesis codes. We begin by introducing the \startrack code, focusing on aspects particularly relevant for SL simulations. In Section \ref{sec:COSMIC}, we describe the \cosmic code with emphasis on its differences compared to \startrack.

\subsection{Star formation history and initial conditions} \label{sec:initial_sfr}

We adopt the model of star formation rate (SFR) and metallicity distributions for the Galactic disk and bulge based on several observational constraints shown in fig. 1-3 of \cite{Olejak2020} with two minor differences. Firstly, we divide disk populations into 10 instead of 11 sub-populations (with no separation for the Thin and Thick disk); secondly, we assume solar metallicity of $Z_{\odot}=0.02$ (instead of $0.014$). We generate 15 stellar populations (10 for the disk and 5 for the bulge) which differ by their metallicity and age. The simulated numbers of systems are scaled to match the total stellar mass of the Milky Way disk and bulge (assumed to be $5.17 \times 10^{10} \msun$ and $0.91 \times 10^{10} \msun$ respectively, \citealt{Licquia15}).

For the initial mass of the primary (the more massive star), we adopt a broken power-law IMF \citep{1993MNRAS.262..545K}, following the standard formula: $\xi(m) = (m/m_0)^{-\alpha_{\rm i}}$, where $\xi(m) dm$ is the number of stars between masses $m$ and $m +dm$. The exponent $\alpha_{i}$ takes values of:
$\alpha_1 = -1.3$ for  $M \in [0.08,0.5]\msun$, \
$\alpha_2 = -2.2$ for  $M \in [0.5,1.0]\msun$, and \
$\alpha_3 = -2.3$ for  $M \in [1.0,150.0]\msun$ \

The mass of the secondary (the less massive star) in the newborn binary is derived from the uniform mass ratio distribution in the range $q \in [0.08/\mathrm{M}_1,1]$. We normalize the stellar mass of the Galactic components, with the assumption of a high binary fraction for massive primary stars (with $M_{\rm ZAMS} > 10 \msun$: $100\%$) and assume that $50\%$ of stars with $M \leq 10 \msun$ are in binary systems (as motivated by observations, e.g. \citealt{Moe2017}).

\subsection{Main assumptions on single and binary stellar astrophysics}\label{sec:assumptions}

Our Galactic population of binary systems hosting compact objects is generated using the \startrack\ population synthesis code \citep{Belczynski2008a, Belczynski2020}. The version of the code used in this paper is the same as that described in Section~2 of \cite{Olejak2022a}. We adopt a strong pulsational-pair-instability/pair-instability supernova which limits the mass of BHs to $\sim45\msun$, as adopted in \cite{Belczynski2016c}. For non-compact accretors, we assume a $50\%$ non-conservative Roche lobe overflow with a fraction of the donor mass ($1-f_{\rm a}$, where $f_{\rm a}$ is the fraction of material transferred from the donor that is accreted onto the compact object) lost from the system, together with the corresponding part of the donor star and orbital angular momentum \citep[see section~3.4 of][]{Belczynski2008a}. We also limit the mass accretion rate onto the non-compact accretor to the Eddington limit. For the stellar wind mass loss, we use formulae based on theoretical predictions of radiation-driven mass loss \citep{Vink2001} with the inclusion of Luminous Blue Variable mass loss \citep{Belczynski2010a}. We adopt a Maxwell distribution of natal kicks, with $\sigma=265$ km/s \citep{Hobbs2005} lowered by fallback \citep{Fryer1204} at NS and BH formation (see section \ref{sec:velocities}). We adopt common envelope (CE) development criteria as described in section~5.2 of \cite{Belczynski2008a}, with the envelope ejection efficiency $\alpha_{\rm CE} = 1.0$. We assume that binary systems with Hertzsprung gap donor stars do not survive the CE phase \citep{Belczynski2008a}, as stars with not well-developed convective envelopes are expected to merge \citep{Klencki2020}.

To calculate stellar remnant masses, we adopt the revised formulae given by \cite{Fryer2022} implemented as described in Section 2.2 of \cite{Olejak2022a}. In contrast to the previously used Rapid and Delayed SN models given by formulae 5 and 6 of \cite{Fryer1204} that may be treated as two extremes, the new formulae allow us to test the impact of assumed convective growth timescales on the remnant masses ($M_{\rm rem}$ in units of $M_{\odot}$). Equation \ref{eq:mrem_max} allows us to calculate remnant masses assuming a smooth relation with a pre-SN carbon-oxygen core mass ($M_{\rm CO}$ in units of $\msun$; in \startrack\ it is the value at the end of a star's core helium burning phase), and by varying different mixing efficiencies ($\fmix$, set by the convection growth timescale):

\begin{equation}\label{eq:mrem_max}
    M_{\rm rem,max} = 1.2 + 0.05 \fmix +\ 0.01 (M_{\rm CO}/\fmix)^2 + e^{\fmix(M_{\rm CO}-M_{\rm crit})}
\end{equation}
\noindent where $M_{\rm crit}=5.75 \msun$ is the assumed critical mass of the carbon-oxygen core for BH formation (the switch from NS to BH formation). In this work, we test three variants of the formula with $\fmix=0.5$, $1.0$, and $4.0$, corresponding to models f05, f10, and f40. Models f05 and f40 correspond to the Delayed and Rapid models from \cite{Fryer1204}, whereas f10 is an intermediate case. Note that the mass of the remnant is calculated using Equation \ref{eq:mrem_max} until some $M_{\rm CO}$ value is reached, which depends on the steepness of the exponent (i.e. the adopted $\fmix$ value). If $M_{\rm rem}$ from Equation \ref{eq:mrem_max} exceeds the value of the total mass of the pre-SN star ($M_{\rm pre-SN}$), then we assume direct collapse of the star to a BH with mass-loss only in the form of neutrinos ($1 \%$ of the pre-SN mass): 

\begin{equation}
    M_{\rm rem} = \min(M_{\rm rem,max}, M_{\rm pre-SN}).
\label{eq:mrem}
\end{equation}

In our simulations, NSs may also form via electron-capture SN \citep{Hiramatsu2021} and are expected to possess relatively low masses and low natal kick velocities \citep{Podsiadlowski2004} in contrast to core-collapse SN NSs, which are expected (and observed, e.g. \citeauthor{Lyne9405}, \citeyear{Lyne9405}) to receive very large kicks as a result of an asymmetric neutrino-driven explosion. Electron-capture SN NSs obtain no additional kicks due to asymmetric explosion (but the orbit may still be affected by the mass loss) and always have a post-SN mass of $1.26\msun$.

We assume that the system doesn't change between SL flares. This should be true in general for the companion star, as the orbital period is far smaller than evolutionary timescales. However, precession resulting from flybys, which can be important also in sparse stellar systems like the Galactic disk \citep{Klencki1708}, may destroy the fragile arrangement of system orbit and the observer.

\subsection{Velocities}\label{sec:velocities}

The centre-of-mass velocities of binary systems play a crucial role in determining their spatial distribution and detectability as SL sources. Systems that receive large natal kicks during SN explosions may be ejected from their birth locations, potentially reducing their detection probability in targeted survey regions. 

We calculate the centre-of-mass velocity ($\vcm$) for each binary using:

\begin{equation}
    \vec{v}_{\rm cm}=\frac{M_\mathrm{a,f} \vec{v}_\mathrm{a,f}+M_\mathrm{b} \vec{v}_\mathrm{b}}{M_\mathrm{a,f} + M_\mathrm{b}}
\end{equation}
\noindent where $\mathrm{M}_\mathrm{a,f}$ and $\vec{v}_\mathrm{a,f}$ are the mass and velocity of the compact object (post-SN), and $\mathrm{M}_\mathrm{b}$ and $\vec{v}_\mathrm{b}$ are the mass and velocity of the companion star.

We estimate each velocity component ($v_x$, $v_y$, $v_z$) using \citep[for derivation, see section~6.3 of][]{Belczynski2008a}:

\begin{equation}
    \vec{v}_{\rm cm} = \vec{v}_{\rm cm,bef} + \alpha \vec{v}_{\rm ka} + \beta \vec{v}_{\rm prea},
\end{equation}
\noindent where $\vec{v}_{\rm cm,bef}$ is the velocity of the binary system before the SN explosion (equal to the rotational velocity around the Galaxy if this is the first SN) and:

\begin{equation}\label{eq:Belczynski08_kick}
\alpha=\frac{M_\mathrm{a,f}}{M_\mathrm{a,f}+M_\mathrm{b}},\quad
\beta=\frac{M_\mathrm{b} (M_\mathrm{a,f}-M_\mathrm{a})}{(M_\mathrm{a}+M_\mathrm{b})(M_\mathrm{a,f}+M_\mathrm{b})}.
\end{equation}

The first term in this equation relates to the asymmetric SN ejection, while the second term represents the Blaauw kick \citep{Blaauw1961} due to mass loss during compact object formation. Here, $\mathrm{M}_{\rm a}$ is the pre-SN mass of the star, $v_{\rm prea,i}$ is the relative velocity of the stars before the SN explosion, and $v_{\rm ka,i}$ is the natal kick velocity of the compact object after the SN (reduced by fallback). $v_{\rm cm,i}$ represents only the additional velocity obtained through evolution, excluding Galactic rotation. We calculate the velocity magnitude as:

\begin{equation}
    \vcm = \sqrt{v_\mathrm{cm,x}^2 + v_\mathrm{cm,y}^2 + v_\mathrm{cm,z}^2}
\end{equation}

\subsection{Cosmic}\label{sec:COSMIC}

Given the substantial uncertainties and assumptions inherent in population synthesis modelling, we performed additional simulations using an independent code to assess the robustness of our results and quantify inter-code systematic uncertainties.

We employed the \emph{Compact Object Synthesis and Monte Carlo Investigation Code} (\cosmic), a binary population synthesis code adapted from the Binary Stellar Evolution code \citep[BSE;][]{Hurley2002} with updated evolution prescriptions and parameters \citep{Breivik2007}. The evolutionary prescriptions are broadly consistent with those implemented in \startrack, though some key differences exist as outlined below.

For the primary star initial mass distribution, we adopt the same broken power-law IMF described in Section \ref{sec:initial_sfr}, with an additional regime for the lowest masses: $\alpha_0 = -0.3$ for $\mathrm{M} \in [0.01, 0.08]\msun$. 

The secondary star mass is sampled from a uniform mass ratio distribution $q \in [M_{2,\mathrm{min}}/M_1, 1]$, where (and unlike with \startrack) $M_{2,\mathrm{min}}$ is constrained such that the pre-main sequence lifetime of the secondary does not exceed the total lifetime of the primary (assuming single-star evolution).

Orbital periods and eccentricities, similarly to \startrack follow the observational distributions from \citet{Sana1207}:
\begin{itemize}
    \item Period distribution: $P \propto (\log P)^{-0.55}$
    \item Eccentricity distribution: $P \propto e^{-0.42}$
\end{itemize}

Stellar winds are computed using theoretical prescriptions for radiation-driven mass loss and Luminous Blue Variable episodes \citep{Vink2001, Vink2005}. Wind velocities and mass accretion rates are consistent with \startrack implementations, with accretion rates onto the NS/BH limited to the Eddington rate.

Mass transfer stability follows the prescriptions of \citet{Hurley2002}, with critical mass ratios consistent with standard BSE formulations. Distinct stability criteria are applied for giant stars following \citet{Hjellming1987}.

Common envelope evolution employs the standard $\alpha\lambda$ formalism with an ejection efficiency parameter $\alpha_{\mathrm{CE}} = 1.0$. The binding energy factor $\lambda$ is calculated using the method of \citet{Pols1995}. Following the \startrack simulations, we adopt a "pessimistic" CE scenario where mergers are inevitable when unstable mass transfer occurs in systems with ill-defined core-envelope boundaries \citep{Klencki2020}.

We implement two SN mechanisms representing the extremes of convective mixing efficiency: the "Delayed" and "Rapid" explosion models of \citet{Fryer1204} (equivalent to models f05 and f40, respectively). 

For remnant mass determination, we apply a maximum mass loss of $0.5\msun$ due to baryonic-to-gravitational mass conversion. This limit applies when the compact object progenitor core mass $\geq 11\msun$, corresponding to complete fallback scenarios where natal kicks are governed by neutrino emission and are therefore negligible.

Natal kicks are drawn from Maxwellian distributions with velocity dispersions of:

\begin{itemize}
    \item $\sigma = 265$ km s$^{-1}$ for Fe core-collapse SN
    \item $\sigma = 20$ km s$^{-1}$ for electron-capture and ultra-stripped SN (progenitor mass range 1.6–2.25 M$_{\odot}$)
\end{itemize}

The above prescriptions follow \citet{Hobbs2005}. While recent observations suggest some BHs may receive significant natal kicks comparable to NSs \citep[e.g.,][]{Atri1911, Fragione2203}, we maintain consistency with \startrack by moderating kicks through fallback effects, resulting in negligible kicks for massive compact object progenitors.

\subsection{Self-lensing description}

The SL calculations in this work follow the methodology established in W21. Simulated binaries are randomly positioned in the Galactic potential as described in \citet{Wiktorowicz2012}, with positions used to calculate apparent magnitudes for each system as observed from Earth (positioned $8.3$ kpc from the Galactic centre and $20$ pc above the Galactic plane). Binary inclinations are sampled uniformly in $\cos i$, ensuring isotropic orbital orientation distributions. Uncertainty estimates were established using bootstrap resampling with 12 iterations with results reported as means and standard errors (1-$\sigma$). For each bootstrapping sample, systems received different Galactic positions and inclinations.

For each system, the lensing amplitude $\musl$ is calculated using the analytic framework of \citet{Witt9408}, which accounts for the finite size of the source star and the impact parameter (see W21 for details). The effective crossing time $\tau_{\rm eff}$ is defined as the duration during which at least a fraction of the companion star's disk falls within the Einstein radius of the compact object, incorporating realistic impact parameters rather than assuming edge-on geometries  (see equation (5) of W21).

A crucial distinction in this work concerns the definition of ``detectable'' SL events. By ``detectable'' we mean systems for which at least one observation can be made during the SL event, given a particular survey. This criterion does not require that the event can be distinguished from other astrophysical phenomena such as stellar flares or outbursts, nor does it account for the detectability of the lensing signal above the photometric noise level of the host star. The analysis presented in subsequent sections considers all systems meeting this basic observational requirement, with more conservative predictions accounting for noise thresholds discussed in Section~\ref{sec:conservative_predictions}.

\section{Results}

\subsection{Intrinsic distribution of masses}

\begin{figure}
\centering
\includegraphics[width=\columnwidth]{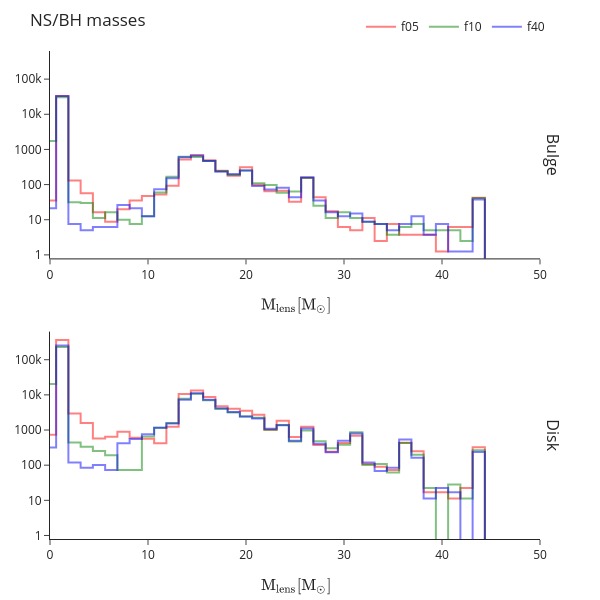}
\caption{Intrinsic mass distribution of Galactic NSs and BHs in binaries with non-compact companions from \startrack. Top panel: bulge sources; bottom panel: disk sources.}
\label{fig:intrinsic_lens}
\end{figure}

\begin{figure}
\centering
\includegraphics[width=\columnwidth]{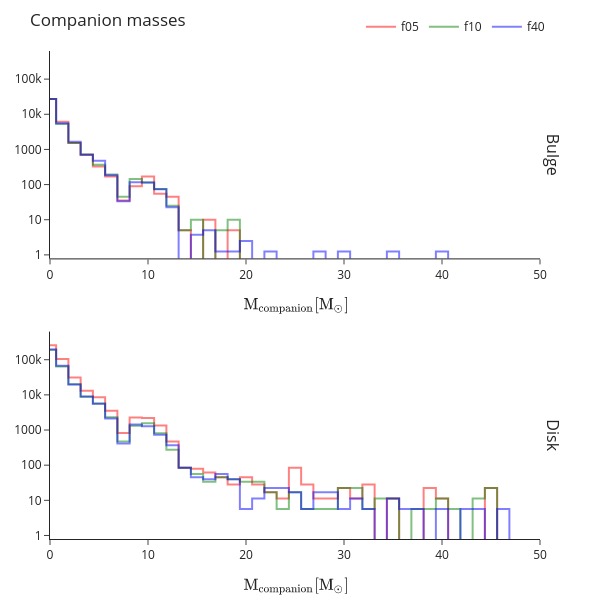}
\caption{Intrinsic mass distribution of non-compact binary companions in Galactic systems from \startrack. Top panel: bulge sources; bottom panel: disk sources.}
\label{fig:intrinsic_companion}
\end{figure}

\begin{figure}
\centering
\includegraphics[width=\columnwidth]{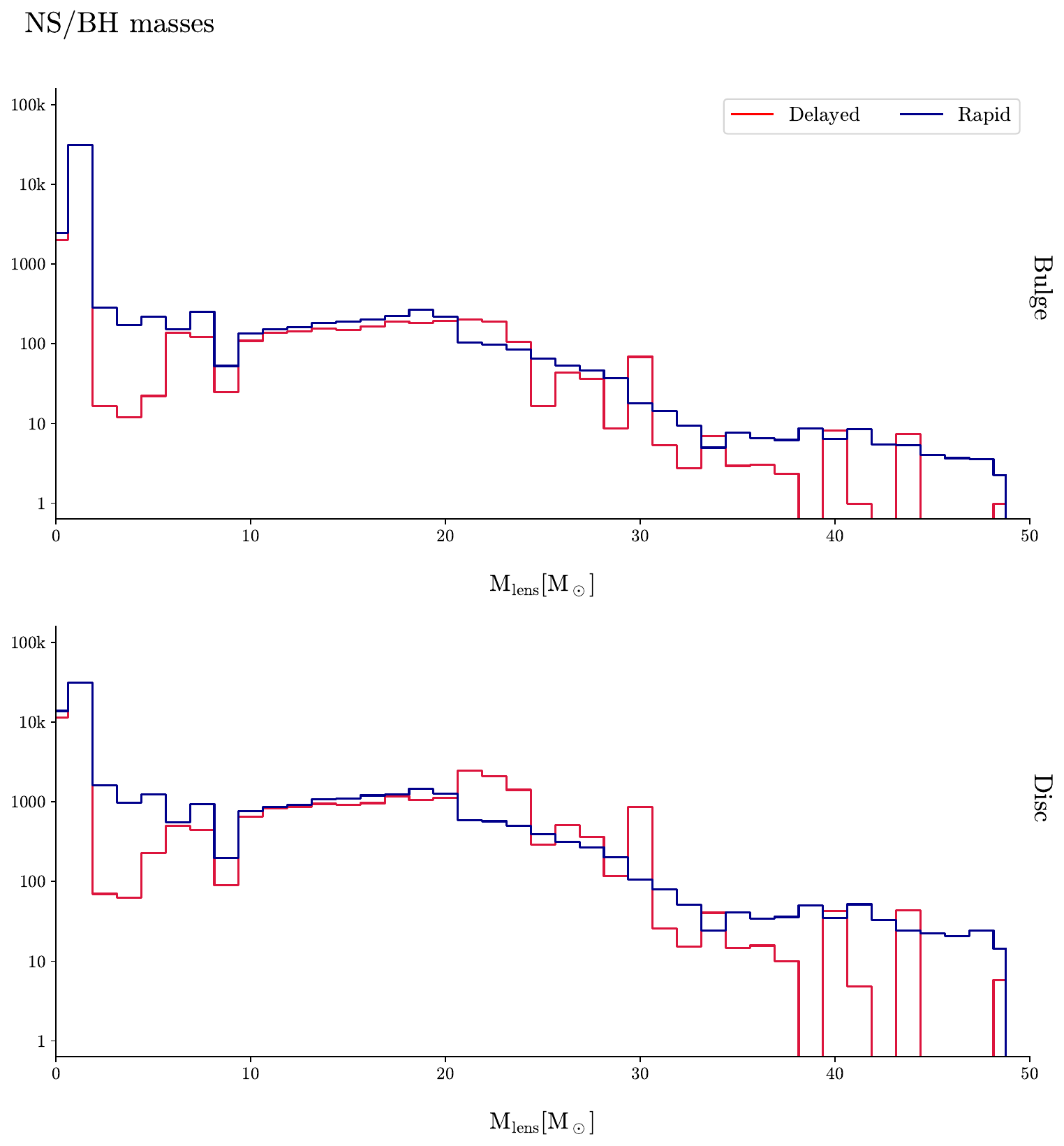}
\caption{Same as Fig~\ref{fig:intrinsic_lens} but from \cosmic.}
\label{fig:intrinsic_lens_cosmic}
\end{figure}

\begin{figure}
\centering
\includegraphics[width=\columnwidth]{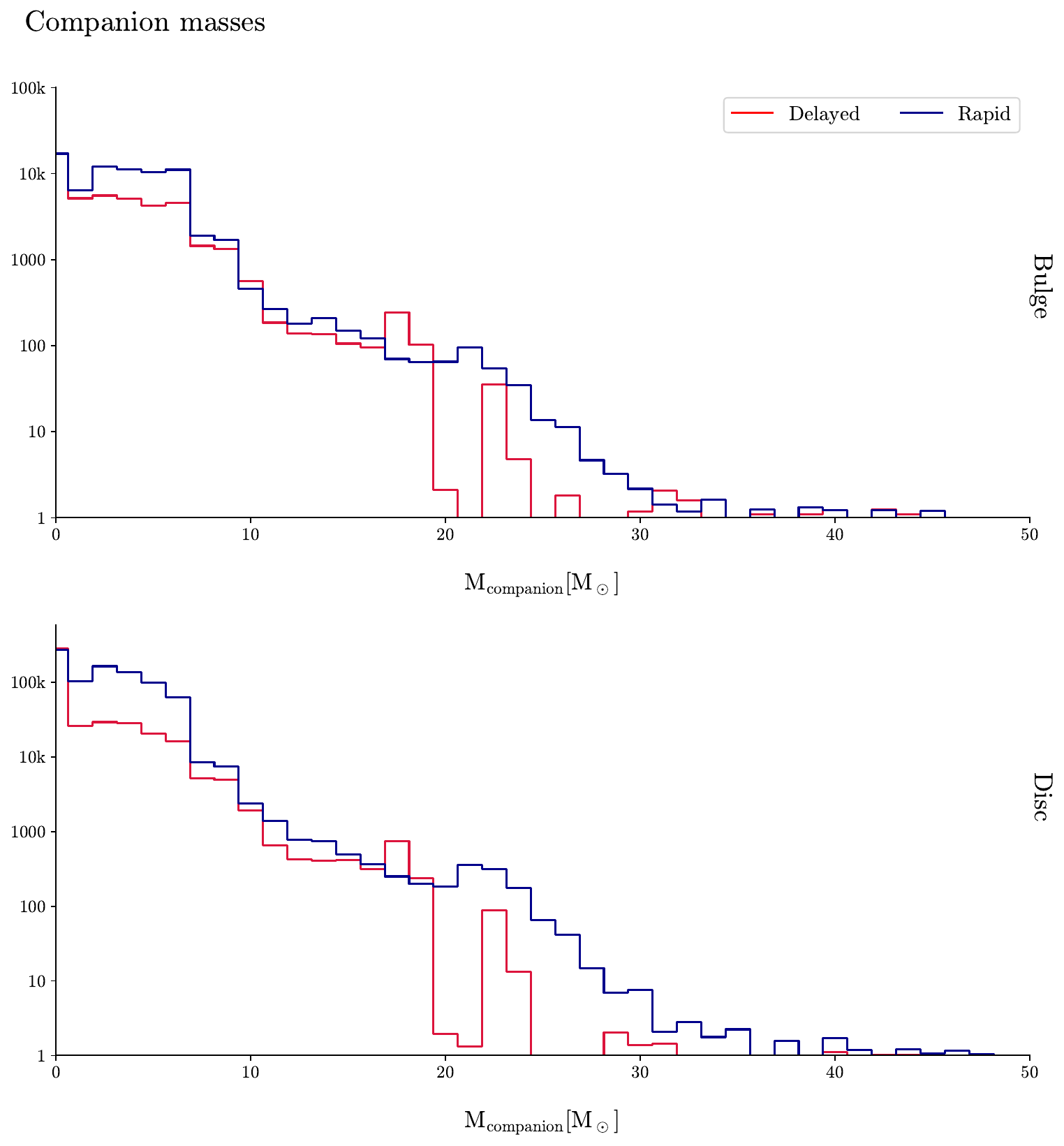}
\caption{Same as Fig~\ref{fig:intrinsic_companion} but from \cosmic.}
\label{fig:intrinsic_companion_cosmic}
\end{figure}

Figs.~\ref{fig:intrinsic_lens} and \ref{fig:intrinsic_companion} show the intrinsic mass distributions of compact objects (the lenses) and their non-compact companions in Galactic binary systems. We present results for three SN models with $\fmix=0.5$, $1.0$, and $4.0$ using \startrack, while Fig.~\ref{fig:intrinsic_lens_cosmic} and \ref{fig:intrinsic_companion_cosmic} show the Delayed and Rapid models for \cosmic.

The higher metallicity in the bulge compared to the disk (Section~\ref{sec:initial_sfr}; \citeauthor{Olejak2020}, \citeyear{Olejak2020}) produces a noticeable decrease in the ratio of BHs to NSs; otherwise, both distributions are similar. The rapid convective growth model ($\fmix=4.0$ or Rapid) creates a prominent mass gap between $2$--$7\msun$. In contrast, the $\fmix=0.5$ model efficiently produces massive NSs and low-mass BHs within this range. The intermediate model ($\fmix=1.0$) produces a gap, but less pronounced than the $\fmix=4.0$ case. Overall, \startrack\ and \cosmic\ distributions are comparable.

The non-compact companion mass distributions are largely independent of the SN model choice, differing only in absolute numbers. The fraction of massive non-compact companions is lower in the bulge due to enhanced stellar wind mass loss at higher metallicity.

\begin{figure}
\includegraphics[width=\columnwidth]{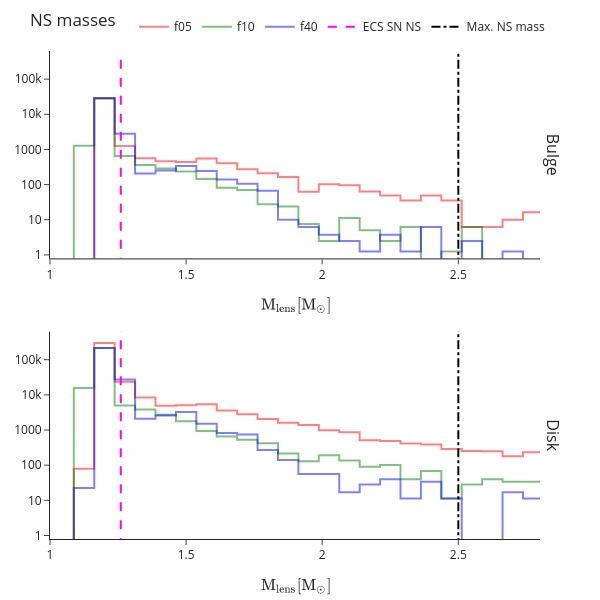}
\caption{Intrinsic mass distribution of Galactic NSs ($\mlens < 2.5\msun$) in binaries with non-compact companions. This shows a zoomed view of the NS population from Fig.~\ref{fig:intrinsic_lens}.}
\label{fig:intrinsic_ns}
\end{figure}

\begin{figure}
\centering
\includegraphics[width=\columnwidth]{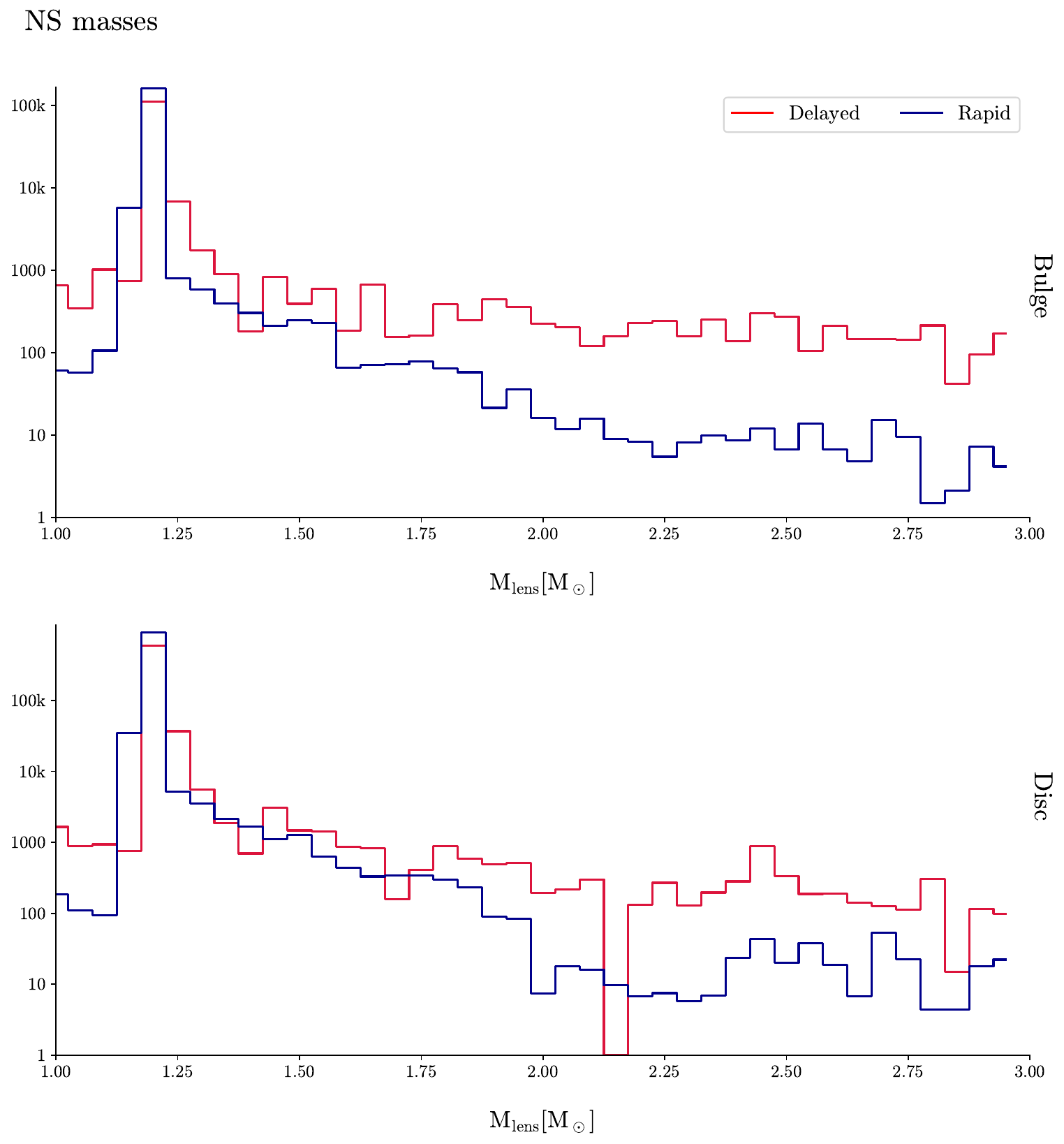}
\caption{Same as Fig.~\ref{fig:intrinsic_ns} but from \cosmic.}
\label{fig:intrinsic_ns_cosmic}
\end{figure}

Fig.~\ref{fig:intrinsic_ns} and \ref{fig:intrinsic_ns_cosmic} focus on the NS mass distributions. The prominent peak at $1.26\msun$ corresponds to NSs formed via electron-capture SN, enhanced by our assumption of zero natal kicks for such events. The $\fmix=0.5$ (Delayed) model differs significantly from other models in producing more massive NSs ($\gtrsim1.5\msun$) and mass gap objects ($\gtrsim2\msun$).

\subsection{Self-Lensing}

\begin{table*}
\begin{tabular}{ll|lllll}
\toprule
mwc & survey & \multicolumn{3}{c}{\startrack} & \multicolumn{2}{c}{\cosmic}\\
 & & f05 & f10 & f40 & Delayed & Rapid\\
\midrule
\multirow[t]{3}{*}{bulge} & LSST & $(1.5\pm 0.0)\times10^{2}$ & $(1.6\pm 0.1)\times10^{2}$ & $(1.6\pm 0.0)\times10^{2}$ & $(6.2)\times10^{2}$ & $(2.8)\times10^{2}$ \\
 & TESS & $(2.1\pm 0.1)\times10^{1}$ & $(2.8\pm 0.2)\times10^{1}$ & $(2.5\pm 0.2)\times10^{1}$ & $(5.5)\times10^{1}$ & $(2.8)\times10^{1}$ \\
 & ZTF & $(3.7\pm 0.1)\times10^{2}$ & $(3.1\pm 0.1)\times10^{2}$ & $(3.5\pm 0.1)\times10^{2}$& $(6.5)\times10^{2}$ & $(3.6)\times10^{2}$\\
\cline{1-7}
\multirow[t]{3}{*}{disk} & LSST & $(2.4\pm 0.1)\times10^{3}$ & $(1.6\pm 0.0)\times10^{3}$ & $(1.6\pm 0.0)\times10^{3}$& $(3.2)\times10^{3}$& $(1.3)\times10^{3}$\\
 & TESS & $(5.9\pm 0.3)\times10^{2}$ & $(3.7\pm 0.3)\times10^{2}$ & $(4.0\pm 0.3)\times10^{2}$& $(7.8)\times10^{2}$ & $(4.3)\times10^{2}$ \\
 & ZTF & $(6.3\pm 0.1)\times10^{3}$ & $(3.8\pm 0.1)\times10^{3}$ & $(3.9\pm 0.1)\times10^{3}$& $(2.7)\times10^{3}$ & $(1.3)\times10^{3}$ \\
\cline{1-7}
\bottomrule
\end{tabular}
\caption{Predicted number of detectable SL events by survey and Galactic component (mwc) for different SN engine models (f05, f10, f40). Both \startrack and \cosmic results are shown. Values represent bootstrap means and standard errors (1-$\sigma$) from 12 iterations. Results do not account for filter effects or signal detectability above noise. \cosmic's Delayed and Rapid models roughly correspond to \startrack's f05 and f40 models, respectively.}
\label{tab:results}
\end{table*}

Table \ref{tab:results} presents our predictions for detectable\footnote{By "detectable" we mean systems for which at least one observation can be made during the SL event. This does not necessarily mean the event can be identified as a SL event, as it may be confused with other types of astrophysical variability or noise.} SL events across different surveys and Galactic components. We used bootstrap resampling with 12 iterations to establish uncertainty estimates, reporting means and standard errors (1-$\sigma$) for each model configuration. These results do not account for filter effects or the detectability of flares above the noise level (see Section~\ref{sec:conservative_predictions}).

Several clear trends emerge from our analysis. ZTF consistently shows the highest detection potential across all models, with predicted yields of $\sim6300$ detectable SL events for the disk using the f05 model. LSST is the second most productive survey with $\sim 3200$ disk events, while TESS shows the lowest yields at $\sim 590$ events. This ranking holds regardless of the binary population model or Galactic component considered, and is consistent with previous findings from W21.

The Galactic disk consistently produces significantly more detectable events than the bulge -- approximately an order of magnitude more across all surveys and models. For example, ZTF predictions range from $\sim 370$ bulge events to $\sim 6300$ disk events for the f05 model, while LSST shows $\sim 150$ bulge events compared to $\sim 3200$ disk events.

Among our SN models, f05 predicts the highest numbers of detectable SL events for disk populations, while the three models yield comparable results for bulge populations. The Delayed \cosmic model generally produces results comparable to our f40 model, though with somewhat more events than f05. This difference is most pronounced for the disk component in LSST and TESS observations, where \cosmic predicts approximately $50\%$ more events in Delayed than \startrack in f05.

The Delayed SN mechanism (f05) consistently produces more observable SL binaries than the Rapid mechanism (f40), particularly in disk populations across both \startrack and \cosmic simulations and primarily in the low $\mlens$ ($\lesssim10\msun$) regime (see Table \ref{tab:results}). This differential response suggests the relative distribution between disk and bulge components could provide valuable constraints on SN mechanisms.

While our analysis identifies potentially detectable events, limited temporal coverage or insufficient orbital repetitions during a survey's operational period may prevent definitive classification (see Section~\ref{sec:coverage_and_recurrence}). However, these candidates remain valuable targets for follow-up radial velocity measurements to independently confirm their nature (McMaster et al., in prep).

\begin{figure*}
\includegraphics[width=\textwidth]{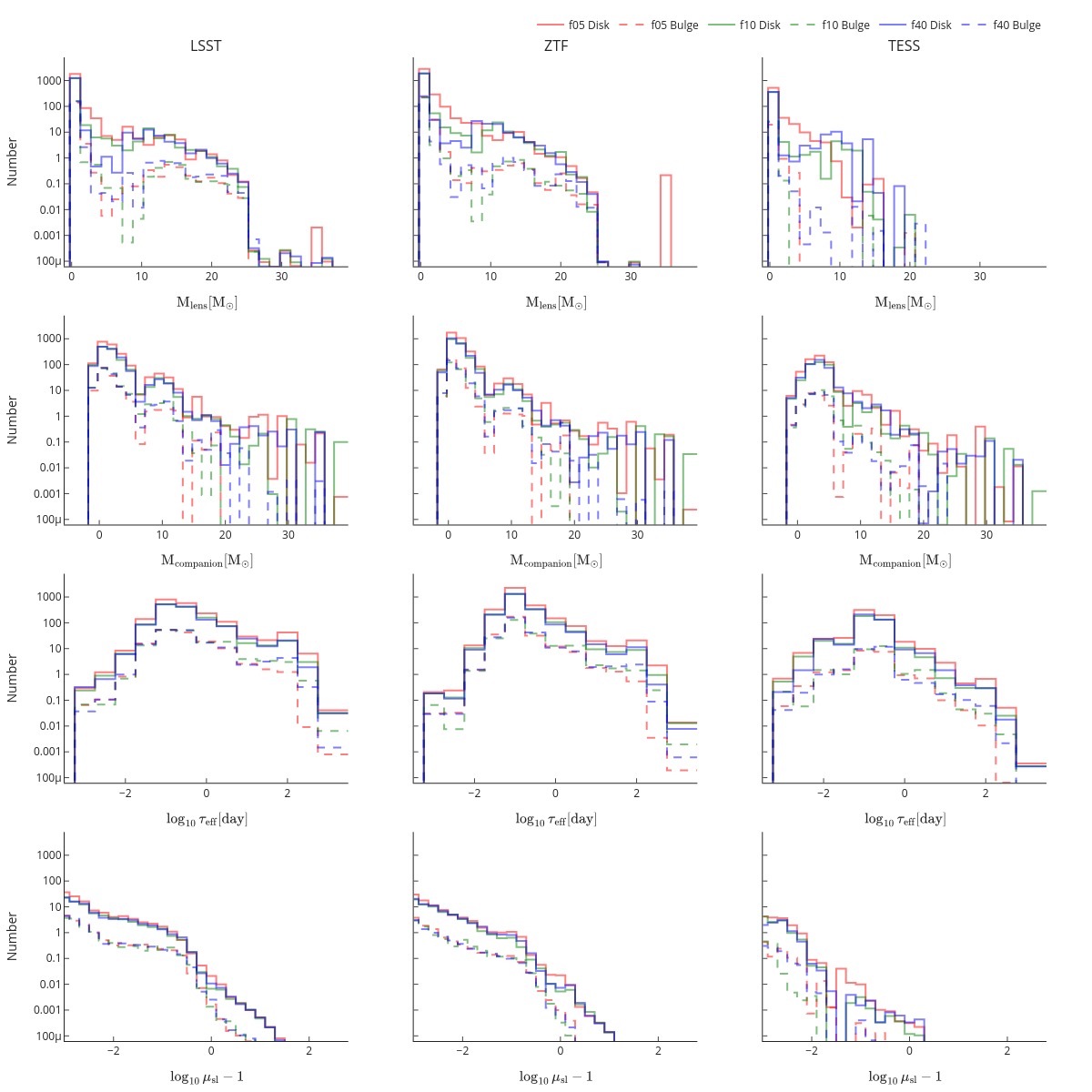}
\caption{Distributions of key parameters for SL binary systems from \startrack simulations for different SN models and survey instruments, separated by Galactic component (disk and bulge). Parameter distributions include lens mass ($\ml$), source mass ($\ms$), effective Einstein crossing time ($\taueff$), and SL magnification ($\musl$). Histogram bars with statistically negligible values ($\ensl\lesssim10^{-3}$) for $\ml$ and $\ms$ are omitted for clarity, and the $\musl$ range is truncated below $10^{-5}$ for clarity (cf. fig. 3 in W21).}
\label{fig:distributions}
\end{figure*}
\begin{figure*}
\centering
\includegraphics[width=\textwidth]{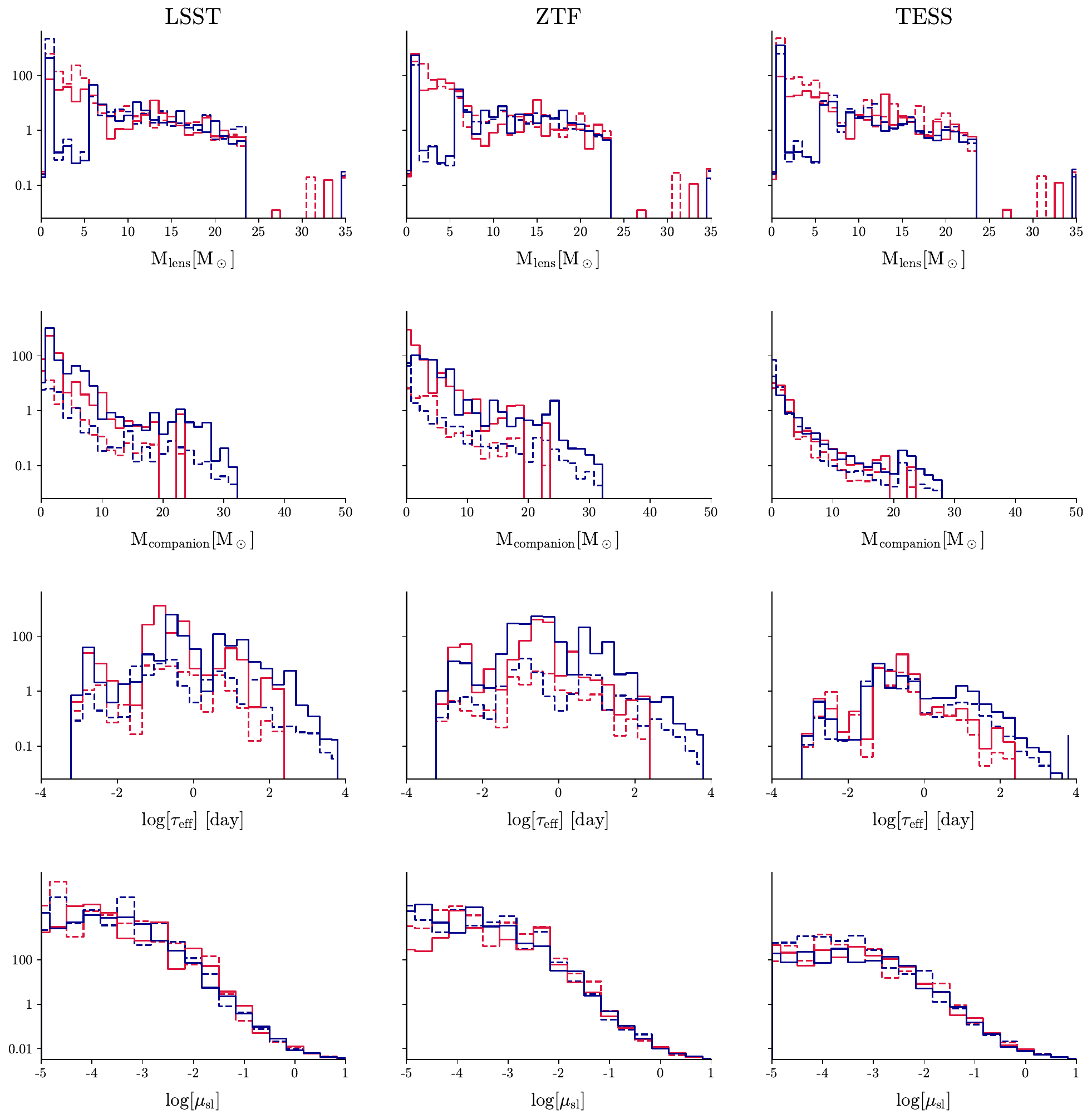}
\caption{Parameter distributions for SL binary systems from \cosmic simulations, equivalent to Fig. \ref{fig:distributions}.}
\label{fig:distributions_cosmic}
\end{figure*}

Fig. \ref{fig:distributions} compares key system parameters from \startrack results across the tested SN models and observing instruments, with separate lines for the Galactic disk and bulge. The $\mlens$ distributions exhibit notable differences between $\fmix$ values, reflecting variations in the underlying intrinsic distributions (see Fig. \ref{fig:intrinsic_lens}). The peak at $\sim35\msun$ in the ZTF's $\mlens$ distribution likely results from a single high-probability system that obtained significant observational weight by chance. 

TESS shows enhanced sensitivity to massive companions ($\gtrsim10\msun$), for which the effective Einstein crossing time ($\taueff$) is characteristically short ($\lesssim10^{-2}$ yr). These close binaries have very short periods that can be effectively sampled only by TESS with its high cadence. Notably, all parameter distributions except $\mlens$ show minimal dependence on the adopted SN model. Similar conclusions can be drawn from the  \cosmic results (Fig.~\ref{fig:distributions_cosmic}).

\subsection{Mass Gap Objects}

\begin{table*}
\centering
Results for the Mass Gap systems\\
\begin{tabular}{lllllll}
\toprule
mwc & instr & \multicolumn{3}{c}{{\tt StarTrack}} & \multicolumn{2}{c}{\cosmic}\\
 & & f05 & f10 & f40 & Delayed & Rapid\\
\midrule
bulge & LSST & $3.9\pm 0.4$ & $1.2\pm 0.4$ & $2.6\pm 1.1$ & 7.6 & $(5.4)\times10^{-1}$\\
bulge & TESS & $10.0\pm 0.7$ & $(1.3\pm 1.8)\times10^{-1}$ & $(3.1\pm 4.6)\times10^{-1}$ & 2.2& $(1.5)\times10^{-1}$\\
bulge & ZTF & $(4.5\pm 0.4)\times10^{1}$ & $5.6\pm 0.8$ & $6.5\pm 1.3$ & 2.1 & $1.9$\\
disk & LSST & $(1.5\pm 0.1)\times10^{2}$ & $(3.1\pm 0.8)\times10^{1}$ & $(1.5\pm 0.5)\times10^{1}$ & $(7.3)\times10^{1}$& $2.9$ \\
disk & TESS & $(5.7\pm 0.7)\times10^{1}$ & $5.5\pm 2.7$ & $1.7\pm 1.4$ &$(1.1)\times10^{1}$ & $4.0$\\
disk & ZTF & $(7.2\pm 0.3)\times10^{2}$ & $(1.3\pm 0.2)\times10^{2}$ & $(6.3\pm 0.9)\times10^{1}$ & $(4.6)\times10^{1}$& $(1.2)\times10^{1}$ \\
\bottomrule
\end{tabular}
\caption{Same as for Table \ref{tab:results} but for mass gap SL systems ($\mlens$ between $2$ and $5\msun$).}
\label{tab:results_mass_gap}
\end{table*}

Table \ref{tab:results_mass_gap} presents predictions for SL systems in the mass gap ($2$--$5\msun$). ZTF exhibits the highest detection rates and provides the strongest discrimination between SN engine models, with \startrack f05 yielding $\sim10$ times more detections than f10 in the disk component. LSST yields substantially fewer mass gap detections compared to ZTF, despite its superior photometric precision, primarily due to its lower observing cadence (assumed to be $4.6$ days versus $1$ day). TESS shows the most limited potential for mass gap discoveries, particularly in the bulge, where rates drop to $<1$ expected detections for several models.

The \cosmic results generally predict lower numbers of expected detections than \startrack, with differences up to a factor of $5$. This disparity reflects the different prescriptions for SN engines, making this lens mass range of $\mlens\lesssim10\msun$ particularly sensitive to the choice of evolutionary codes. Although \startrack's f05 and f40 models are counterparts of \cosmic's Delayed and Rapid models respectively, the prescriptions differ, which may account for the observed differences in estimated numbers of detections. Furthermore, this range of mass is particularly susceptible to evolutionary effects such as SN explosions and natal kicks, which may be sources of additional differences between the models.

\begin{figure*}
\includegraphics[width=\textwidth]{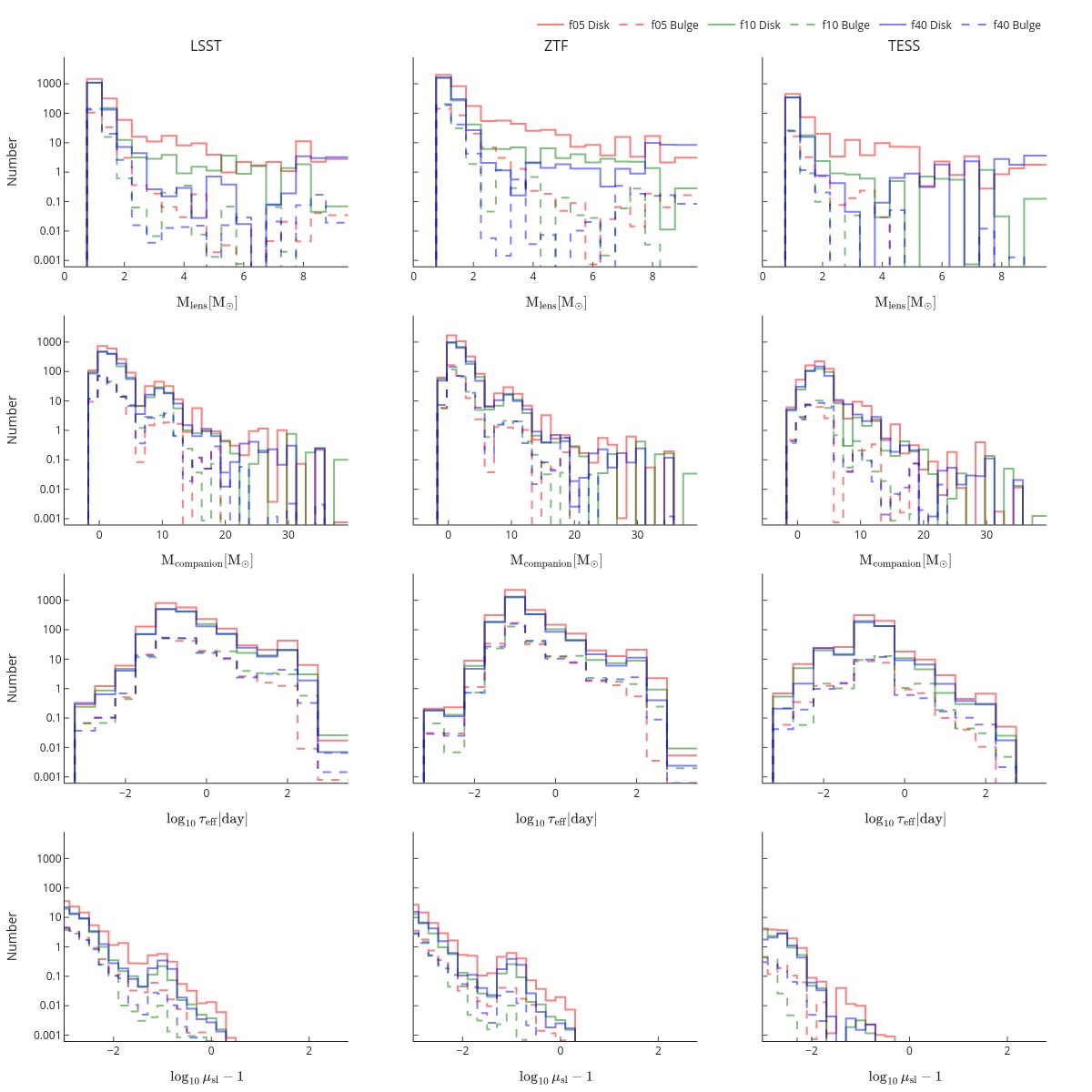}
\caption{Parameter distributions for SL binaries with lens masses in the "mass gap" region. The plotted lens mass ($\ml$) range is restricted to $0$--$10\msun$ to highlight differences in the mass gap ($2$--$5\msun$). All other aspects are identical to Fig. \ref{fig:distributions}.}
\label{fig:distributions_massgap}
\end{figure*}
\begin{figure*}
\centering
\includegraphics[width=\textwidth]{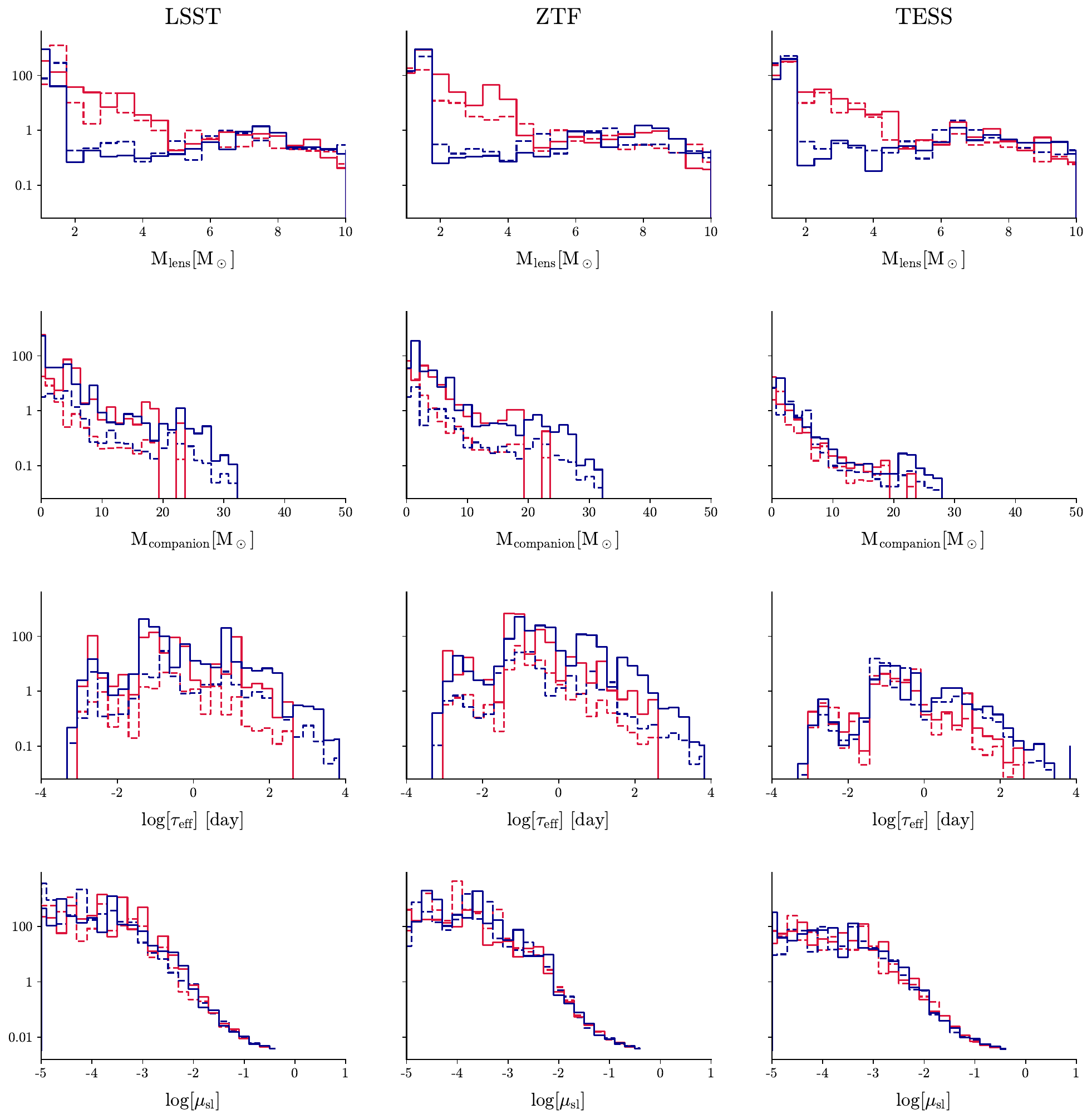}
\caption{Equivalent of Fig. \ref{fig:distributions_massgap} from \cosmic.}
\label{fig:distributions_massgap_cosmic}
\end{figure*}

Fig. \ref{fig:distributions_massgap} presents the same parameter distributions as Fig. \ref{fig:distributions}, but focuses exclusively on systems with lens masses ($\mlens$) between 0 and $10\msun$. The distributions exhibit similar overall trends to those observed in the complete population. However, notable differences between models emerge primarily in the $\mlens$ distributions (this distinction is also evident in the \cosmic data shown in Fig. \ref{fig:distributions_massgap_cosmic}). Compared to the entire population, the subset of lenses with low $\taueff$ values is more pronounced in this mass range. Beyond this effect, the primary observable difference remains the enhanced detection frequency characteristic of the f05 model.

\subsection{Coverage and recurrence}\label{sec:coverage_and_recurrence}

SL events in binary systems present unique observational challenges and opportunities. The detectability and characterization of these events depend critically on the observational coverage of individual events and their recurrence over time. In this section, we analyse the implications of our model predictions for various survey instruments, focusing on two key aspects: the number of observations per SL event and the frequency of recurring events for each system. We also introduce a combined metric to assess the overall observability of SL systems across different surveys. This analysis provides crucial insights into the practical limitations and potential strategies for detecting and studying SL events in large-scale astronomical surveys.
    
\begin{table*}
\begin{tabular}{lllcccccc}
\toprule
 &  & mwc: & \multicolumn{3}{c}{disk} & \multicolumn{3}{c}{bulge} \\
 &  & $\ncov$: & 0 & 1 & 10 & 0 & 1 & 10 \\
instr & model & $\nrec$ &  &  &  &  &  &  \\
\midrule
\multirow[t]{9}{*}{LSST} & \multirow[t]{3}{*}{f05} & 0 & $(2.4\pm 0.1)\times10^{3}$ & $(2.0\pm 0.2)\times10^{2}$ & $(8.2\pm 0.5)\times10^{1}$ & $(1.5\pm 0.0)\times10^{2}$ & $(1.1\pm 0.1)\times10^{1}$ & $2.6\pm 0.8$ \\
 &  & 1 & $(2.2\pm 0.1)\times10^{3}$ & $(1.6\pm 0.2)\times10^{2}$ & $(5.3\pm 0.6)\times10^{1}$ & $(1.3\pm 0.0)\times10^{2}$ & $9.9\pm 1.3$ & $1.8\pm 0.8$ \\
 &  & 10 & $(1.8\pm 0.1)\times10^{3}$ & $(9.0\pm 1.6)\times10^{1}$ & $0.0\pm 0.0$ & $(1.0\pm 0.0)\times10^{2}$ & $7.6\pm 1.8$ & $0.0\pm 0.0$ \\
\cline{2-9}
 & \multirow[t]{3}{*}{f10} & 0 & $(1.6\pm 0.0)\times10^{3}$ & $(1.3\pm 0.2)\times10^{2}$ & $(4.1\pm 0.6)\times10^{1}$ & $(1.6\pm 0.1)\times10^{2}$ & $(2.1\pm 0.2)\times10^{1}$ & $6.8\pm 1.1$ \\
 &  & 1 & $(1.5\pm 0.0)\times10^{3}$ & $(1.1\pm 0.2)\times10^{2}$ & $(2.7\pm 0.6)\times10^{1}$ & $(1.5\pm 0.1)\times10^{2}$ & $(1.9\pm 0.2)\times10^{1}$ & $5.1\pm 1.3$ \\
 &  & 10 & $(1.2\pm 0.0)\times10^{3}$ & $(7.2\pm 1.4)\times10^{1}$ & $0.0\pm 0.0$ & $(1.2\pm 0.1)\times10^{2}$ & $(1.2\pm 0.2)\times10^{1}$ & $0.0\pm 0.0$ \\
\cline{2-9}
 & \multirow[t]{3}{*}{f40} & 0 & $(1.6\pm 0.0)\times10^{3}$ & $(1.1\pm 0.1)\times10^{2}$ & $(3.9\pm 1.0)\times10^{1}$ & $(1.6\pm 0.0)\times10^{2}$ & $(1.6\pm 0.2)\times10^{1}$ & $7.5\pm 1.2$ \\
 &  & 1 & $(1.5\pm 0.0)\times10^{3}$ & $(9.8\pm 1.1)\times10^{1}$ & $(2.5\pm 0.7)\times10^{1}$ & $(1.5\pm 0.0)\times10^{2}$ & $(1.5\pm 0.2)\times10^{1}$ & $6.4\pm 1.2$ \\
 &  & 10 & $(1.2\pm 0.0)\times10^{3}$ & $(5.9\pm 0.9)\times10^{1}$ & $0.0\pm 0.0$ & $(1.1\pm 0.0)\times10^{2}$ & $7.4\pm 1.2$ & $0.0\pm 0.0$ \\
\cline{1-9} \cline{2-9}
\multirow[t]{9}{*}{TESS} & \multirow[t]{3}{*}{f05} & 0 & $(5.9\pm 0.3)\times10^{2}$ & $(5.9\pm 0.3)\times10^{2}$ & $(5.8\pm 0.3)\times10^{2}$ & $(2.1\pm 0.1)\times10^{1}$ & $(2.1\pm 0.1)\times10^{1}$ & $(2.0\pm 0.1)\times10^{1}$ \\
 &  & 1 & $(5.4\pm 0.3)\times10^{2}$ & $(5.4\pm 0.3)\times10^{2}$ & $(5.3\pm 0.3)\times10^{2}$ & $(1.8\pm 0.1)\times10^{1}$ & $(1.8\pm 0.1)\times10^{1}$ & $(1.7\pm 0.1)\times10^{1}$ \\
 &  & 10 & $(2.0\pm 0.2)\times10^{2}$ & $(2.0\pm 0.2)\times10^{2}$ & $(1.9\pm 0.2)\times10^{2}$ & $5.4\pm 1.5$ & $5.4\pm 1.5$ & $5.0\pm 1.4$ \\
\cline{2-9}
 & \multirow[t]{3}{*}{f10} & 0 & $(3.7\pm 0.3)\times10^{2}$ & $(3.7\pm 0.3)\times10^{2}$ & $(3.6\pm 0.3)\times10^{2}$ & $(2.8\pm 0.2)\times10^{1}$ & $(2.8\pm 0.2)\times10^{1}$ & $(2.8\pm 0.2)\times10^{1}$ \\
 &  & 1 & $(3.4\pm 0.3)\times10^{2}$ & $(3.4\pm 0.3)\times10^{2}$ & $(3.3\pm 0.3)\times10^{2}$ & $(2.4\pm 0.2)\times10^{1}$ & $(2.4\pm 0.2)\times10^{1}$ & $(2.4\pm 0.2)\times10^{1}$ \\
 &  & 10 & $(1.2\pm 0.1)\times10^{2}$ & $(1.2\pm 0.1)\times10^{2}$ & $(1.1\pm 0.1)\times10^{2}$ & $7.3\pm 1.1$ & $7.3\pm 1.1$ & $7.1\pm 1.1$ \\
\cline{2-9}
 & \multirow[t]{3}{*}{f40} & 0 & $(4.0\pm 0.3)\times10^{2}$ & $(4.0\pm 0.3)\times10^{2}$ & $(3.9\pm 0.3)\times10^{2}$ & $(2.5\pm 0.2)\times10^{1}$ & $(2.5\pm 0.2)\times10^{1}$ & $(2.5\pm 0.2)\times10^{1}$ \\
 &  & 1 & $(3.6\pm 0.3)\times10^{2}$ & $(3.6\pm 0.3)\times10^{2}$ & $(3.5\pm 0.3)\times10^{2}$ & $(2.3\pm 0.2)\times10^{1}$ & $(2.3\pm 0.2)\times10^{1}$ & $(2.2\pm 0.2)\times10^{1}$ \\
 &  & 10 & $(1.4\pm 0.2)\times10^{2}$ & $(1.4\pm 0.2)\times10^{2}$ & $(1.3\pm 0.3)\times10^{2}$ & $6.2\pm 1.2$ & $6.2\pm 1.2$ & $5.9\pm 1.3$ \\
\cline{1-9} \cline{2-9}
\multirow[t]{9}{*}{ZTF} & \multirow[t]{3}{*}{f05} & 0 & $(6.3\pm 0.1)\times10^{3}$ & $(5.3\pm 0.2)\times10^{2}$ & $(1.0\pm 0.1)\times10^{2}$ & $(3.7\pm 0.1)\times10^{2}$ & $(3.1\pm 0.2)\times10^{1}$ & $4.8\pm 1.0$ \\
 &  & 1 & $(6.1\pm 0.1)\times10^{3}$ & $(4.9\pm 0.2)\times10^{2}$ & $(7.3\pm 0.8)\times10^{1}$ & $(3.6\pm 0.1)\times10^{2}$ & $(3.0\pm 0.2)\times10^{1}$ & $3.5\pm 1.1$ \\
 &  & 10 & $(6.0\pm 0.1)\times10^{3}$ & $(3.8\pm 0.2)\times10^{2}$ & $(1.5\pm 0.4)\times10^{1}$ & $(3.4\pm 0.1)\times10^{2}$ & $(2.5\pm 0.2)\times10^{1}$ & $2.2\pm 0.9$ \\
\cline{2-9}
 & \multirow[t]{3}{*}{f10} & 0 & $(3.8\pm 0.1)\times10^{3}$ & $(3.3\pm 0.3)\times10^{2}$ & $(5.1\pm 0.7)\times10^{1}$ & $(3.1\pm 0.1)\times10^{2}$ & $(4.4\pm 0.4)\times10^{1}$ & $7.9\pm 1.3$ \\
 &  & 1 & $(3.7\pm 0.1)\times10^{3}$ & $(3.2\pm 0.3)\times10^{2}$ & $(3.4\pm 0.6)\times10^{1}$ & $(3.0\pm 0.1)\times10^{2}$ & $(4.0\pm 0.4)\times10^{1}$ & $4.0\pm 1.0$ \\
 &  & 10 & $(3.6\pm 0.1)\times10^{3}$ & $(2.6\pm 0.2)\times10^{2}$ & $9.6\pm 3.1$ & $(2.9\pm 0.1)\times10^{2}$ & $(3.3\pm 0.4)\times10^{1}$ & $1.1\pm 0.9$ \\
\cline{2-9}
 & \multirow[t]{3}{*}{f40} & 0 & $(3.9\pm 0.1)\times10^{3}$ & $(3.1\pm 0.2)\times10^{2}$ & $(6.1\pm 1.2)\times10^{1}$ & $(3.5\pm 0.1)\times10^{2}$ & $(3.5\pm 0.4)\times10^{1}$ & $9.0\pm 1.4$ \\
 &  & 1 & $(3.8\pm 0.1)\times10^{3}$ & $(2.9\pm 0.2)\times10^{2}$ & $(4.7\pm 1.0)\times10^{1}$ & $(3.4\pm 0.1)\times10^{2}$ & $(3.2\pm 0.4)\times10^{1}$ & $6.3\pm 1.3$ \\
 &  & 10 & $(3.7\pm 0.1)\times10^{3}$ & $(2.2\pm 0.2)\times10^{2}$ & $(1.1\pm 0.4)\times10^{1}$ & $(3.3\pm 0.1)\times10^{2}$ & $(2.4\pm 0.3)\times10^{1}$ & $1.8\pm 0.6$ \\
\cline{1-9} \cline{2-9}
\bottomrule
\end{tabular}
\caption{Model predictions showing the number of detectable SL events for different thresholds of coverage ($\ncov$) and recurrence ($\nrec$). Results are presented for each survey, model and Galactic component.}
\label{table:coverage_and_recurrence}
\end{table*}

Table~\ref{table:coverage_and_recurrence} presents the results of our simulations for different surveys, obtained by imposing observational cuts (limiting magnitudes, cadence, etc.) on the simulated data (see W21 for details). We introduce two important parameters: coverage ($\ncov$) and recurrence ($\nrec$).

Coverage ($\ncov$) represents the minimum number of observational datapoints covering a single SL event. Higher coverage increases the likelihood of identifying the event with SL and estimating a corresponding magnification and Einstein crossing time. Our results show that the predicted number of sources drops rapidly with increasing $\ncov$ (Table \ref{table:coverage_and_recurrence}), indicating that most SL events will have limited coverage and may be difficult to distinguish from other types of outbursts.
    
Recurrence ($\nrec$) represents the number of recurring SL events for a single source expected during the survey. Recurrence is a key differentiating factor between SL and microlensing events, the latter being one-time occurrences. The main factor limiting $\nrec$ is survey duration. Long-term surveys like LSST or ZTF are more likely to observe multiple events from a single system, whereas short-duration surveys like TESS (effective duration < 2 months) will observe only a few systems with multiple events. In reality, $\nrec$ may be affected by factors that change orbital alignment or system position, such as flybys or precession.

Both coverage and recurrence are negatively correlated with the predicted number of SL detections. However, higher values for both can improve identification through increased observations during peak magnification. We introduce a parameter $\ntot$ describing the total number of observations of lensing events per source during the entire survey duration:
\begin{equation}
    \ntot\ =\ \ncov\times\nrec.
\end{equation} 

\begin{table*}
\begin{tabular}{lllcccc}
\toprule
 &  & $\ntot$ & 0 & 10 & 20 & 50 \\
mwc & instr & model &  &  &  &  \\
\midrule
\multirow[t]{9}{*}{bulge} & \multirow[t]{3}{*}{LSST} & f05 & $(1.5\pm 0.0)\times10^{2}$ & $(1.0\pm 0.0)\times10^{2}$ & $(7.6\pm 0.4)\times10^{1}$ & $(3.0\pm 0.3)\times10^{1}$ \\
 &  & f10 & $(1.6\pm 0.1)\times10^{2}$ & $(1.2\pm 0.1)\times10^{2}$ & $(9.6\pm 0.6)\times10^{1}$ & $(4.4\pm 0.4)\times10^{1}$ \\
 &  & f40 & $(1.6\pm 0.0)\times10^{2}$ & $(1.2\pm 0.0)\times10^{2}$ & $(8.8\pm 0.4)\times10^{1}$ & $(4.1\pm 0.2)\times10^{1}$ \\
\cline{2-7}
 & \multirow[t]{3}{*}{TESS} & f05 & $(2.1\pm 0.1)\times10^{1}$ & $(2.1\pm 0.1)\times10^{1}$ & $(2.1\pm 0.1)\times10^{1}$ & $(2.1\pm 0.1)\times10^{1}$ \\
 &  & f10 & $(2.8\pm 0.2)\times10^{1}$ & $(2.8\pm 0.2)\times10^{1}$ & $(2.8\pm 0.2)\times10^{1}$ & $(2.8\pm 0.2)\times10^{1}$ \\
 &  & f40 & $(2.5\pm 0.2)\times10^{1}$ & $(2.5\pm 0.2)\times10^{1}$ & $(2.5\pm 0.2)\times10^{1}$ & $(2.5\pm 0.2)\times10^{1}$ \\
\cline{2-7}
 & \multirow[t]{3}{*}{ZTF} & f05 & $(3.7\pm 0.1)\times10^{2}$ & $(3.4\pm 0.1)\times10^{2}$ & $(3.3\pm 0.1)\times10^{2}$ & $(2.7\pm 0.1)\times10^{2}$ \\
 &  & f10 & $(3.1\pm 0.1)\times10^{2}$ & $(3.0\pm 0.1)\times10^{2}$ & $(2.8\pm 0.1)\times10^{2}$ & $(2.3\pm 0.1)\times10^{2}$ \\
 &  & f40 & $(3.5\pm 0.1)\times10^{2}$ & $(3.4\pm 0.1)\times10^{2}$ & $(3.2\pm 0.1)\times10^{2}$ & $(2.7\pm 0.1)\times10^{2}$ \\
\cline{1-7} \cline{2-7}
\multirow[t]{9}{*}{disk} & \multirow[t]{3}{*}{LSST} & f05 & $(2.4\pm 0.1)\times10^{3}$ & $(1.9\pm 0.1)\times10^{3}$ & $(1.3\pm 0.1)\times10^{3}$ & $(5.1\pm 0.4)\times10^{2}$ \\
 &  & f10 & $(1.6\pm 0.0)\times10^{3}$ & $(1.2\pm 0.0)\times10^{3}$ & $(8.7\pm 0.3)\times10^{2}$ & $(3.5\pm 0.2)\times10^{2}$ \\
 &  & f40 & $(1.6\pm 0.0)\times10^{3}$ & $(1.2\pm 0.0)\times10^{3}$ & $(8.7\pm 0.4)\times10^{2}$ & $(3.3\pm 0.3)\times10^{2}$ \\
\cline{2-7}
 & \multirow[t]{3}{*}{TESS} & f05 & $(5.9\pm 0.3)\times10^{2}$ & $(5.9\pm 0.3)\times10^{2}$ & $(5.9\pm 0.3)\times10^{2}$ & $(5.9\pm 0.3)\times10^{2}$ \\
 &  & f10 & $(3.7\pm 0.3)\times10^{2}$ & $(3.7\pm 0.3)\times10^{2}$ & $(3.7\pm 0.3)\times10^{2}$ & $(3.7\pm 0.3)\times10^{2}$ \\
 &  & f40 & $(4.0\pm 0.3)\times10^{2}$ & $(4.0\pm 0.3)\times10^{2}$ & $(4.0\pm 0.3)\times10^{2}$ & $(4.0\pm 0.3)\times10^{2}$ \\
\cline{2-7}
 & \multirow[t]{3}{*}{ZTF} & f05 & $(6.3\pm 0.1)\times10^{3}$ & $(6.1\pm 0.1)\times10^{3}$ & $(5.8\pm 0.1)\times10^{3}$ & $(4.9\pm 0.1)\times10^{3}$ \\
 &  & f10 & $(3.8\pm 0.1)\times10^{3}$ & $(3.7\pm 0.1)\times10^{3}$ & $(3.5\pm 0.1)\times10^{3}$ & $(2.9\pm 0.1)\times10^{3}$ \\
 &  & f40 & $(3.9\pm 0.1)\times10^{3}$ & $(3.8\pm 0.1)\times10^{3}$ & $(3.6\pm 0.1)\times10^{3}$ & $(3.0\pm 0.1)\times10^{3}$ \\
\cline{1-7} \cline{2-7}
\bottomrule
\end{tabular}
\caption{Model predictions for the total number of observations ($\ntot$) of SL events during the entire survey duration. Results are shown for different survey instruments and SN models.}
\label{table:total_points}
\end{table*}

Table \ref{table:total_points} summarizes our results for different thresholds of $\ntot$. LSST is capable of detecting a few tens of well-covered ($\ntot > 50$) SL systems, which should allow for better inference of curve features and tighter constraints on system parameters \citep[e.g.,][]{Kruse1404}. For LSST and ZTF, we observe a statistically significant difference between predictions for different SN models, which becomes more pronounced with higher $\ntot$ thresholds.

\subsection{Clustering of Self-Lensing Systems}

\begin{figure*}
\includegraphics[width=\textwidth]{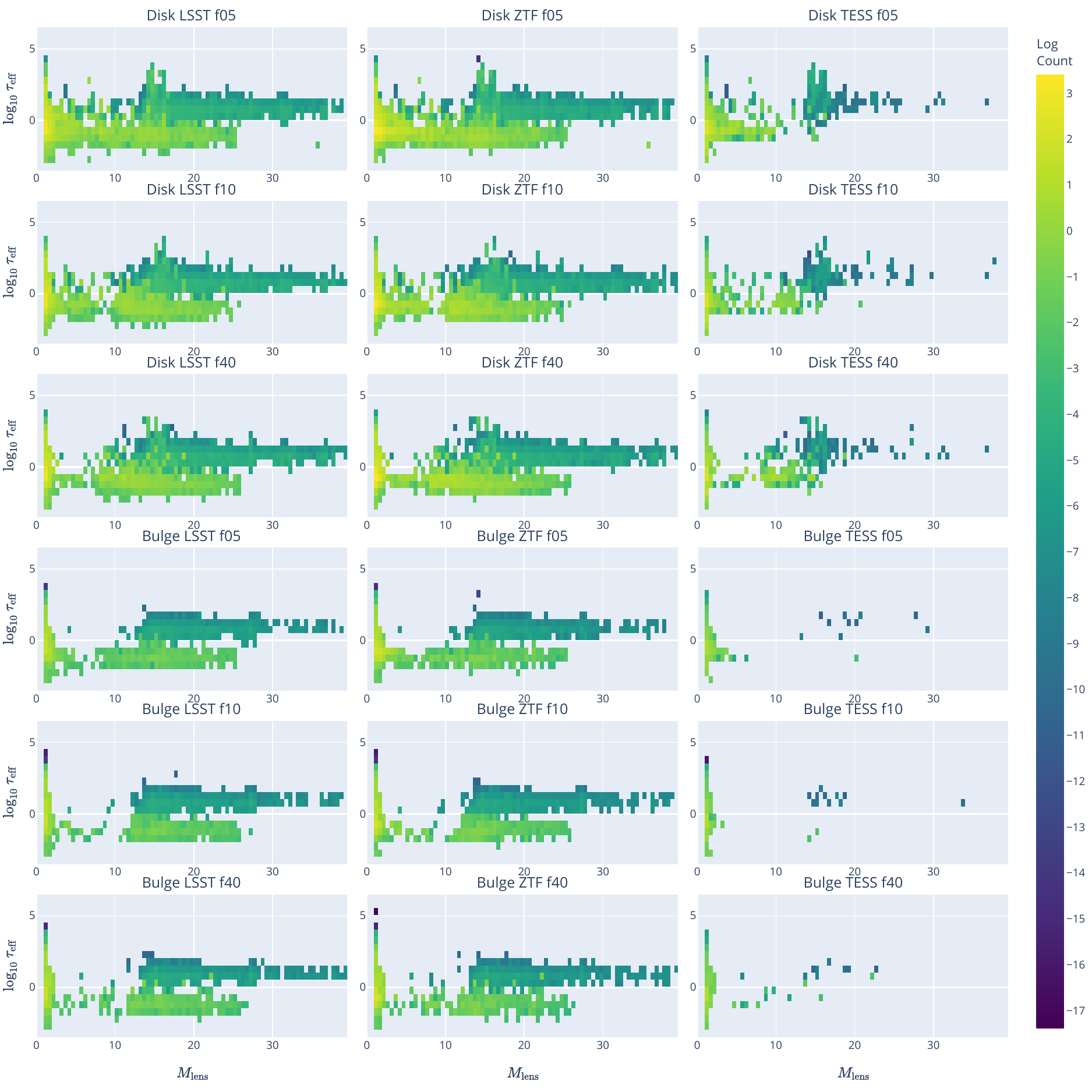}
\caption{Relationship between effective Einstein crossing time ($\taueff$) and lens mass ($\mlens$) across Galactic components, surveys, and SN engine models. The color scale is consistent across all panels, revealing three distinct populations of SL systems. The mass gap region ($\mlens \in [2,5]$) shows the highest population density in the f05 model and lowest in the f40 model.
}
\label{fig:log_tau_eff-Ml}
\end{figure*}
Fig.~\ref{fig:log_tau_eff-Ml} illustrates the relationship between $\taueff$ and $\mlens$. All SN models exhibit similar patterns, with the mass gap region ($\mlens \in [2,5]$) most prominently populated in the f05 model and least populated in the f40 model. While TESS results follow comparable trends, the lower predicted number of SL sources makes these patterns less distinct.

Three distinct groups of objects are evident in Fig.~\ref{fig:log_tau_eff-Ml}:
\begin{itemize}
\item \gbhs: Higher-mass, short-duration systems ($\mlens > 2$ and $\taueff < 1$ day)
\item \gbhl: Higher-mass, long-duration systems ($\mlens > 2$ and $\taueff \geq 1$ day)
\item \gns: Lower-mass systems ($\mlens \leq 2$)
\end{itemize}

\begin{table*}
\begin{tabular}{lrrrrrrrrr}
\toprule
group & $\langle\taueff\rangle$ & $\langle\ml\rangle$ & $\langle\ms\rangle$ & $\langle\porb\rangle$ & $\langle\ncov\rangle$ & $\langle\nrec\rangle$ & $\langle\vcm\rangle$ & $\langle\ntot\rangle$ & $\langle\muslmax\rangle$ \\
\midrule
\gbhs & $0.16$ & $4.09$ & $1.50$ & $2.08$ & $1.00$ & $75.20$ & $118.60$ & $106.60$ & $13.59$ \\
\gbhl & $1.86$ & $3.25$ & $0.52$ & $24.47$ & $1.17$ & $74.10$ & $98.86$ & $90.30$ & $200.92$ \\
\gns & $0.22$ & $1.29$ & $2.12$ & $5.35$ & $1.00$ & $58.20$ & $28.13$ & $80.81$ & $350.91$ \\
\bottomrule
\end{tabular}
\caption{Median properties of the SL sources in different groups visible in Fig.~\ref{fig:log_tau_eff-Ml} for each survey and Galactic component. All values are medians weighted by detection probability. See text for group definitions.
Properties include effective Einstein crossing time ($\taueff$ [day]), lens mass ($\ml [\msun]$), source mass ($\ms [\msun]$), orbital period ($\porb$ [day]), coverage ($\ncov$), recurrence ($\nrec$), centre-of-mass velocity ($\vcm [\kms]$), total coverage ($\ntot$), and highest magnification ($\muslmax$).}
\label{tab:log_tau_eff-ML_groups}
\end{table*}
Table \ref{tab:log_tau_eff-ML_groups} summarizes the weighted median properties of each population, revealing distinctive characteristics beyond just $\mlens$ and $\taueff$:
\begin{itemize}
    \item The \gns population features more massive source stars compared to the \gbhs and \gbhl groups.
    \item \gns systems exhibit the highest magnifications ($\langle\muslmax\rangle$), though the majority ($\sim80\%$) show modest magnifications ($\musl \lesssim 2$).
    \item Although \gns can have much higher centre-of-mass velocities (up to $\sim500\kms$), both \gbhs and \gbhl populations have higher median values ($\langle\vcm\rangle$), suggesting greater resilience to natal kicks due to their more massive compact objects.
    \item All populations display similar typical number of datapoints per one SL event (coverage; $\ncov$) and typical number of recurring SL events per source during survey duration (recurrence; $\nrec$). See Section~\ref{sec:coverage_and_recurrence}.
    \item The \gbhl population, characterized by longer $\taueff$, the consequence of longer orbital periods ($\porb$), shows substantially higher magnification ($\musl$) than the \gbhs population.
\end{itemize}

While most systems have relatively low recurrence rates, the few systems with higher $\nrec$ (see Section~\ref{sec:coverage_and_recurrence}) significantly influence the weighted median due to their enhanced detectability. The elevated typical total coverage count ($\ntot$) directly results from the high typical recurrence values.

The systems with the highest magnifications typically harbour low mass main-sequence source stars ($\ms\lesssim0.5\msun$) on very wide orbits ($\porb>10$ kyr).  Such systems will have very small probability of observation and their configuration will probably be unstable due to dynamical interactions \citep{Klencki1708}.

\subsection{Velocity in detail}\label{sec:vel}

\begin{figure}
    \centering
    \includegraphics[width=\linewidth]{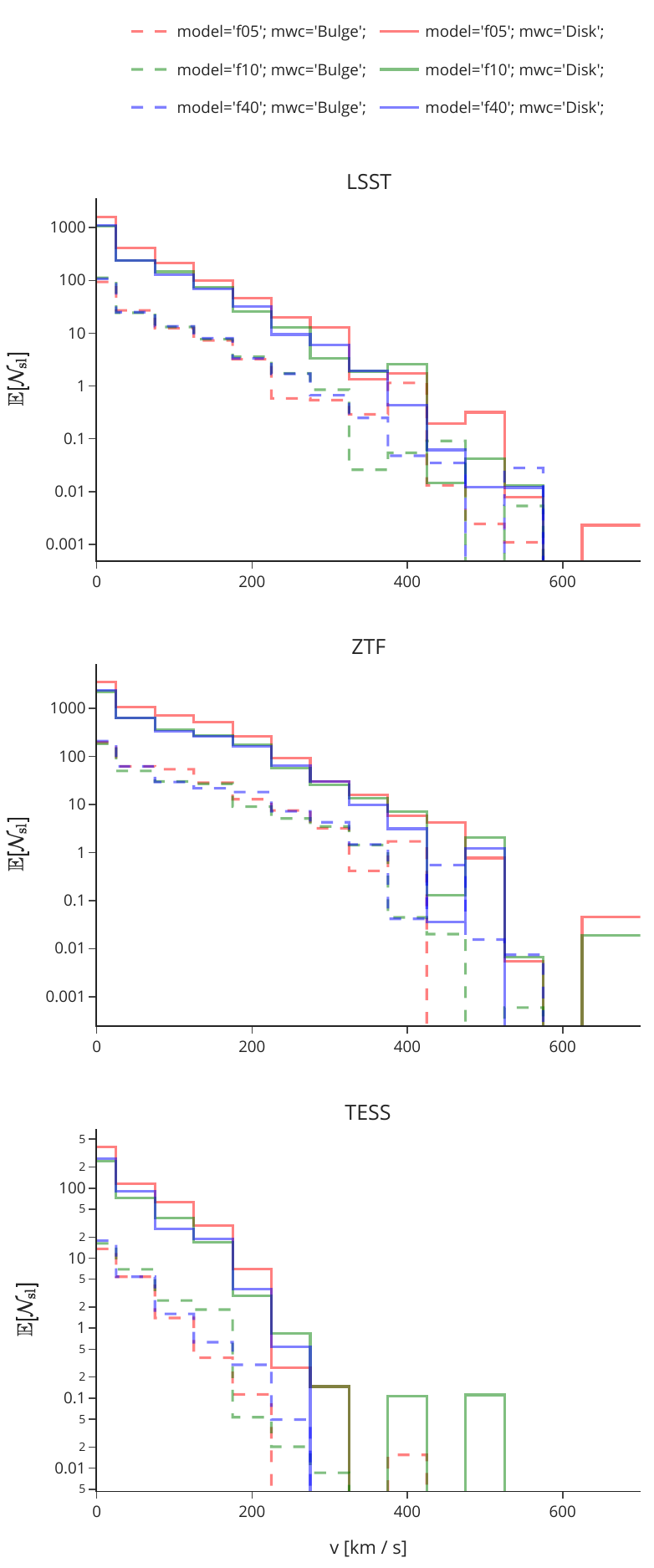}
    \caption{Center-of-mass velocity ($\vcm$) distributions for SL binaries for different SN engines and surveys, separated by Galactic component (disk and bulge).} 
    \label{fig:v_by_instr}
\end{figure}

Fig.~\ref{fig:v_by_instr} presents the centre-of-mass velocity ($\vcm$) distributions of SL sources for different SN engines and surveys, separated by Galactic component. The distributions show minimal variation between SN engines for the same survey. Notably, despite the lower predicted number of systems in the bulge compared to the disk, their velocity distributions remain similar. Any differences observed in the high-velocity tails ($\vcm > 300$ km/s) are statistically insignificant due to limited sample sizes. The distributions are dominated by low-velocity systems ($\lesssim20$ km/s), primarily representing BHs with significant or complete fallback, and NSs receiving low kicks from the assumed, underlying Maxwell distribution.

\begin{figure}
    \centering
    \includegraphics[width=\linewidth]{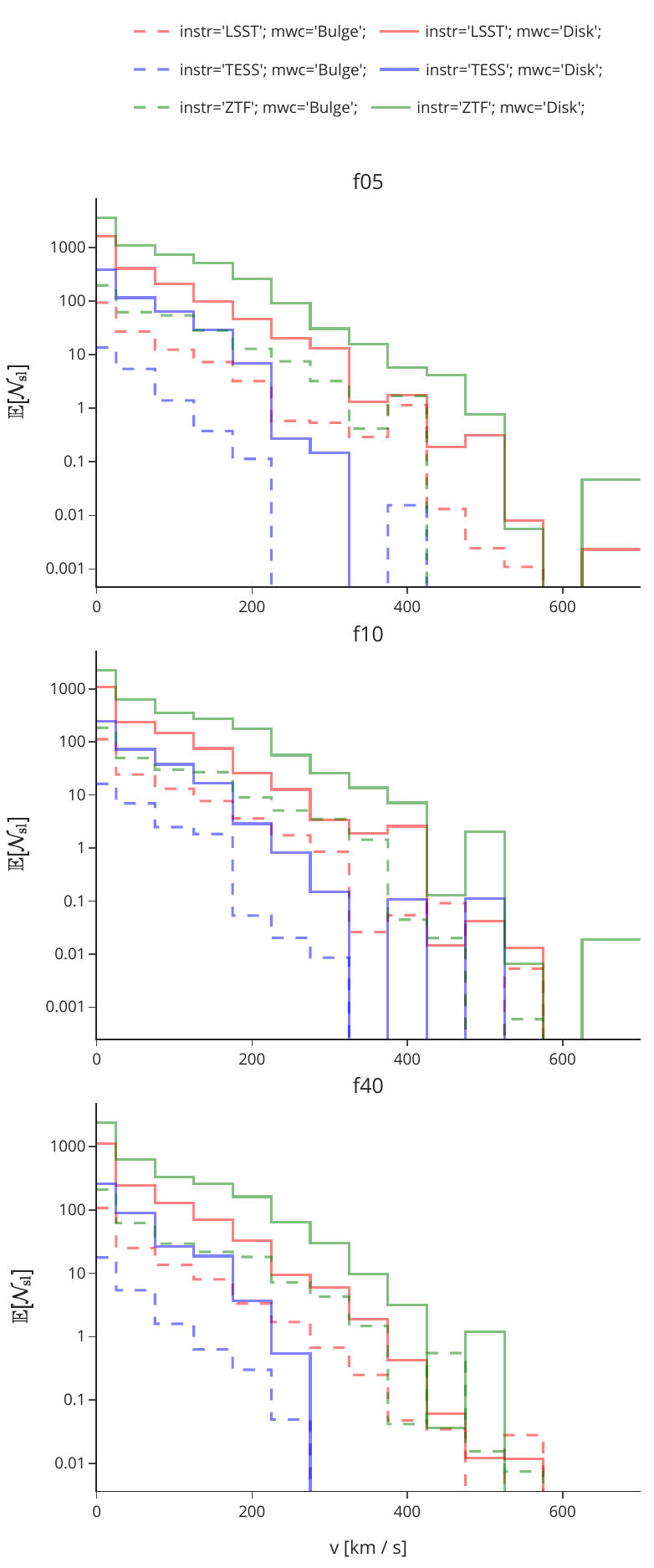}
    \caption{Center-of-mass velocity ($\vcm$) distributions comparing different surveys (LSST, ZTF, and TESS) for each SN engine and Galactic component.}
    \label{fig:v_by_model}
\end{figure}
Fig.~\ref{fig:v_by_model} illustrates the differences in velocity distributions between surveys. While LSST and ZTF show similar distributions (with ZTF predicting higher absolute numbers of sources), TESS exhibits a distinct cut-off around $300$ km/s with few systems beyond this threshold. This cut-off occurs because rapidly moving systems typically form wide, eccentric binaries that are rarely captured by TESS's limited exposure time per sky region. As expected, these survey-dependent characteristics remain consistent across all SN models and Galactic components.

The predominance of low-velocity systems ($\lesssim20$ km/s) stems from a fundamental selection effect: binaries experiencing smaller natal kicks are more likely to remain gravitationally bound. Consequently, more massive SN remnants — which undergo significant fallback or direct collapse with minimal momentum kicks — are preferentially preserved as binary systems. Similarly, NSs formed through electron-capture SN, which receive no natal kicks in our simulations, have higher survival rates in binaries compared to NSs formed through core-collapse SN, which typically receive kicks exceeding $100$ km/s (see Section \ref{sec:assumptions}). Therefore, the velocity distribution reflects the physical expectation that most detectable SL events will occur in binaries that have experienced minimal disruption from SN kicks.

\begin{figure*}
\includegraphics[width=\textwidth]{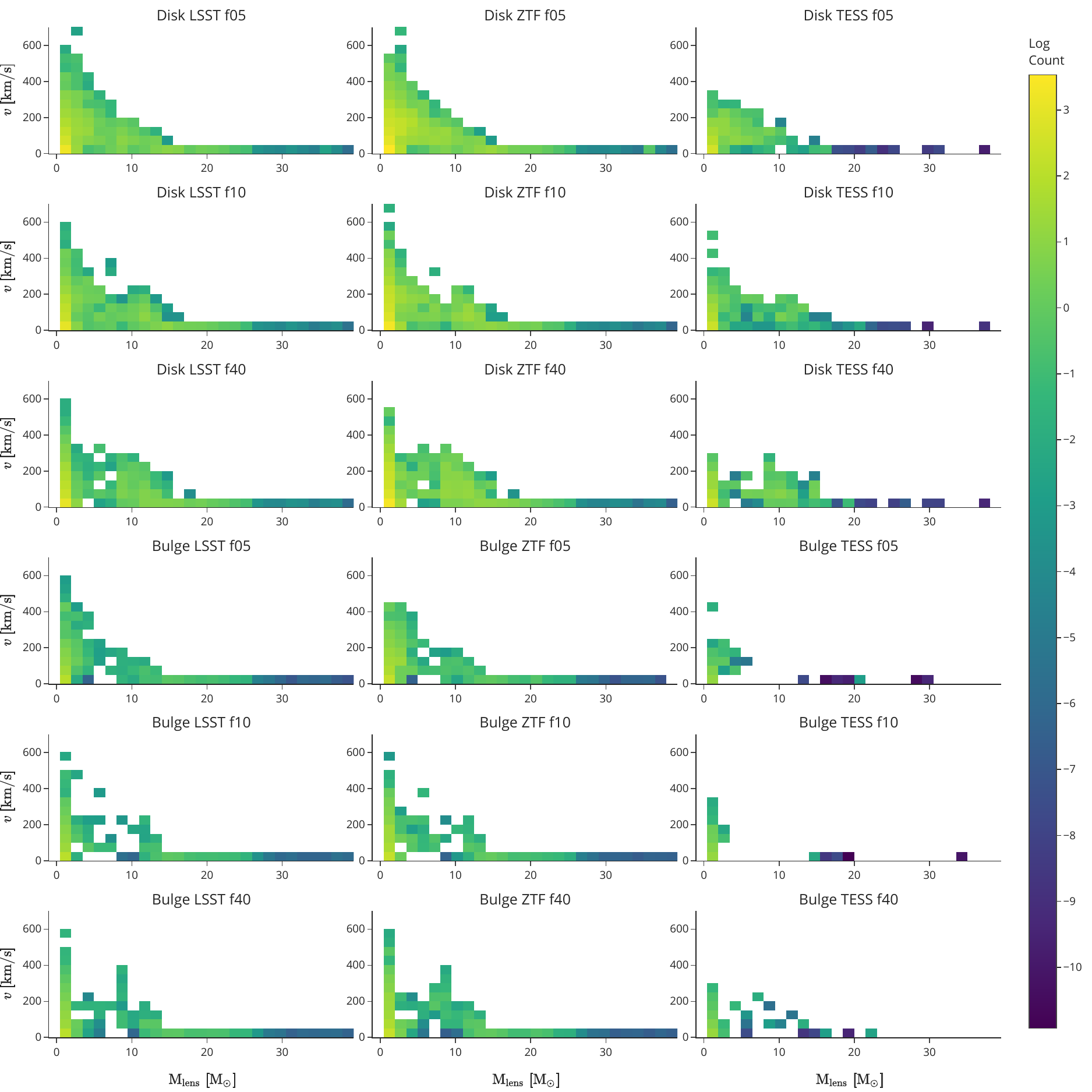}
\caption{Relationship between lens mass ($\ml$) and center-of-mass velocity ($\vcm$) across all models. The columns allow comparison between different SN engines, while rows differentiate between surveys (LSST, ZTF, and TESS) and Galactic components (disk and bulge).}
\label{fig:Ml_vs_v}
\end{figure*}
Fig.~\ref{fig:Ml_vs_v} illustrates the relationship between lens mass ($\mlens$) and centre-of-mass velocity ($\vcm$) across all models. Our findings are consistent with the anticorrelation between BH mass and peculiar velocity reported by \citet{Gandhi1905}. While they identified this trend in Galactic BH X-ray binaries using Gaia DR2 data (finding a slope of $-10.6^{+4.6}_{-6.2}$ km s$^{-1}$ $\msun^{-1}$ with $99.9\%$ confidence in the anticorrelation), our models extend this relationship across a broader mass spectrum showing similar trend. We similarly observe that systems with massive lenses ($\gtrsim 40\msun$) exhibit minimal velocities ($\lesssim 1$ km/s), while the highest velocities ($\gtrsim 100$ km/s) correspond to lower-mass lenses ($\lesssim 10\msun$), predominantly NSs. Notably, \citet{Gandhi1905} observe peculiar velocities of $50$ -- $150$ km/s for several systems in this mass range, further strengthening the inverse relationship between compact object mass and system velocity. Interestingly, SL binaries containing mass gap lenses ($\sim2$--$5\msun$) display higher average velocities compared to other SL systems, which results from higher natal kicks from core-collapse SNe and lower fallback.

\section{Conservative detection limits}\label{sec:conservative_predictions}

The values presented throughout this paper are considered raw detection estimates because they do not account for the specific limitations of filter bandwidth or flare detectability caused by noise. In this section, we address this by applying additional constraints.

\subsection{Apparent Magnitude Calculations}\label{sec:apparent_magnitude}

To determine the apparent magnitudes of stars in SL binary systems, we employed a synthetic photometry approach using blackbody radiation profiles and instrument-specific filter responses. Filter transmission curves were obtained from the Filter Profile Service\footnote{https://svo2.cab.inta-csic.es/svo/theory/fps/} for the relevant instruments.

For each source star, we assumed blackbody radiation characterized by the stellar effective temperature $T_{\rm eff}$. The spectral energy distribution was calculated using the Planck function:
\begin{equation}
B_\lambda(T_{\rm eff}) = \frac{2hc^2}{\lambda^5} \frac{1}{\exp\left(\frac{hc}{\lambda k T_{\rm eff}}\right) - 1}
\end{equation}
where $h$ is Planck's constant, $c$ is the speed of light, $k$ is Boltzmann's constant, and $\lambda$ is the wavelength.

The observed flux density at Earth was computed by scaling the blackbody flux by the geometric dilution factor:
\begin{equation}
F_\lambda = \pi B_\lambda(T_{\rm eff}) \left(\frac{R_\star}{D}\right)^2
\end{equation}
where $R_\star$ is the stellar radius and $D$ is the distance to the system.

The photon count rate through each filter was calculated by integrating the product of the flux density and filter response function $R(\lambda)$ over wavelength:
\begin{equation}
C = \int \frac{F_\lambda \lambda R(\lambda)}{hc} \, d\lambda
\end{equation}

The apparent magnitude was then computed as:
\begin{equation}\label{eq:apparent_magnitude}
\msourcesl = -2.5 \log_{10}\left(\frac{C}{C_0}\right)
\end{equation}
\noindent where $C_0$ is the zero counts reference point calculated as:
\begin{equation}
C_{0} = \int \frac{F_{\nu,0}}{\lambda h} \cdot R(\lambda) \, d\lambda
\end{equation}
\noindent where $F_{\nu,0}=3631\times10^{-26}$ [W m$^{-2}$ Hz$^{-1}$].

For SL events, the magnified apparent magnitude was computed by multiplying the flux by the magnification factor $\musl$ before conversion to magnitude:
\begin{equation}
\msourcesl = -2.5 \log_{10}\left(\frac{\musl C}{C_0}\right)
\end{equation}

\subsection{Flare Detection Criteria}

In the simplest (i.e. in the absence of stellar noise - see Ward et al. in prep for more detailed simulations), the detectability of SL flares depends on both the system visibility and the photometric precision of the observing instrument. We applied two fundamental criteria for flare detection:

\begin{description}

    \item[System Visibility:] The source star must be detectable above the instrument's limiting magnitude:
\begin{equation}
\msource < \mlim
\end{equation}
where $\mlim$ is the limiting magnitude for each survey in each single observation. This condition was already included in W21 predictions (see their table 1) and in the raw results throughout the paper. We note this is a rather conservative limitation and can be improved, for example by stacking observations.

    \item[Flare Detectability:] The magnification during the SL event:
    
\begin{equation}
\msource - \msourcesl = 2.5\log_{10}\musl/\sigmanoise > X
\end{equation}

\noindent must produce a photometric signal exceeding X $\sigma$ above the quiescent stellar flux. In the above, $m_{\rm source,SL}$ is the apparent magnitude during the SL event (calculated using the magnification factor $\musl$ as described in Section \ref{sec:apparent_magnitude}), and $\sigmanoise$ is the survey-specific photometric uncertainty at that magnitude.

\end{description}

The photometric noise levels vary significantly between instruments; for LSST, we adopted a uniform photometric precision of $\sigmanoise = 10$ mmag based on the survey specifications\footnote{\url{https://rubinobservatory.org/for-scientists/rubin-101/key-numbers}}; for ZTF, we assumed a conservative estimate of $\sigmanoise = 25$ mmag \citep{Masci1901}; for TESS, the photometric uncertainty depends on the source brightness \citep{Sullivan1508,Oelkers1809}. We utilized the empirical noise model ($\sigmanoise(T)$) from \citet{Oelkers1809}\footnote{Noise model data obtained from Luke Bouma's repository: \url{https://github.com/lgbouma/tnm/blob/master/results/selected_noise_model_good_coords.csv} (R. Oelkers, private communication). Note that the noise is in units of ppm/$10^6$ in the file.}, which provides noise estimates in parts-per-million (ppm) as a function of apparent magnitude. The conversion from ppm to mmag and from 60 to 1-minute exposures is:
\begin{equation}
\begin{split}
\sigma_{\text{noise,TESS}}(T) \text{ [mmag]} &= \frac{\sigma_{\text{noise}}(T) \text{ [ppm]}}{1000} \times \sqrt{\frac{1\,\text{h}}{1\,\text{min}}} \\[1em]
&\approx 0.0077 \times \sigma_{\text{noise}}(T) \text{ [ppm]},
\end{split}
\end{equation}
\noindent where $T$ is TESS magnitude.

Additionally, we apply a noise correction for all instruments to account for multiple detections as 
\begin{equation}
    \sigma_{\text{noise, corr}}=\frac{1}{\sqrt{\ntot}}\sigmanoise,
\end{equation}
\noindent where $\ntot$ is the total number of flare datapoints during the entire survey (including multiple detections of the same flare and repeating events; see Section \ref{sec:coverage_and_recurrence}).

\subsection{Conservative predictions}

\begin{table*}
\begin{tabular}{llllll}
\toprule
instr & mwc & model & $n_\text{sl, filter only}$ & $n_\mathrm{sl,2\sigma}$ & $n_\mathrm{sl,3\sigma}$ \\

\midrule
\multicolumn{6}{c}{\startrack}\\
\hline
\multirow[t]{9}{*}{LSST} & \multirow[t]{3}{*}{bulge} & f05 & $3.0\pm 0.2$ & $2.2\pm 0.2$ & $2.0\pm 0.2$ \\
 &  & f10 & $3.1\pm 0.2$ & $2.2\pm 0.1$ & $1.9\pm 0.1$ \\
 &  & f40 & $3.1\pm 0.1$ & $2.3\pm 0.1$ & $2.1\pm 0.1$ \\
\cline{2-6}
 & \multirow[t]{3}{*}{disk} & f05 & $(3.8\pm 0.2)\times10^{1}$ & $(2.6\pm 0.2)\times10^{1}$ & $(2.0\pm 0.1)\times10^{1}$ \\
 &  & f10 & $(3.1\pm 0.2)\times10^{1}$ & $(2.1\pm 0.1)\times10^{1}$ & $(1.7\pm 0.1)\times10^{1}$ \\
 &  & f40 & $(3.1\pm 0.2)\times10^{1}$ & $(2.2\pm 0.2)\times10^{1}$ & $(1.8\pm 0.1)\times10^{1}$ \\
\cline{1-6} \cline{2-6}
\multirow[t]{9}{*}{TESS} & \multirow[t]{3}{*}{bulge} & f05 & $(6.2\pm 5.5)\times10^{-1}$ & $(1.7\pm 1.6)\times10^{-1}$ & $(3.6\pm 7.0)\times10^{-2}$ \\
 &  & f10 & $(3.4\pm 2.3)\times10^{-1}$ & $(1.2\pm 1.2)\times10^{-1}$ & $(2.4\pm 2.3)\times10^{-2}$ \\
 &  & f40 & $(4.0\pm 1.7)\times10^{-1}$ & $(1.1\pm 0.9)\times10^{-1}$ & $(4.3\pm 6.5)\times10^{-2}$ \\
\cline{2-6}
 & \multirow[t]{3}{*}{disk} & f05 & $(1.3\pm 0.5)\times10^{1}$ & $3.1\pm 1.3$ & $1.4\pm 0.8$ \\
 &  & f10 & $8.0\pm 2.9$ & $2.1\pm 0.9$ & $(9.2\pm 8.4)\times10^{-1}$ \\
 &  & f40 & $(1.4\pm 0.3)\times10^{1}$ & $4.0\pm 1.4$ & $1.6\pm 0.9$ \\
\cline{1-6} \cline{2-6}
\multirow[t]{9}{*}{ZTF} & \multirow[t]{3}{*}{bulge} & f05 & $2.4\pm 0.2$ & $1.5\pm 0.1$ & $(9.8\pm 1.0)\times10^{-1}$ \\
 &  & f10 & $2.9\pm 0.3$ & $1.7\pm 0.2$ & $10.0\pm 0.1$ \\
 &  & f40 & $3.0\pm 0.3$ & $1.6\pm 0.2$ & $(9.2\pm 0.9)\times10^{-1}$ \\
\cline{2-6}
 & \multirow[t]{3}{*}{disk} & f05 & $(4.0\pm 0.3)\times10^{1}$ & $(1.9\pm 0.2)\times10^{1}$ & $(1.2\pm 0.1)\times10^{1}$ \\
 &  & f10 & $(3.3\pm 0.2)\times10^{1}$ & $(1.5\pm 0.1)\times10^{1}$ & $7.3\pm 0.3$ \\
 &  & f40 & $(3.6\pm 0.3)\times10^{1}$ & $(1.7\pm 0.1)\times10^{1}$ & $9.4\pm 0.7$ \\
\midrule
\multicolumn{6}{c}{\cosmic}\\
\hline
\multirow[t]{9}{*}{LSST} & \multirow[t]{3}{*}{bulge} & Delayed & 3.2\,$\pm$\,0.4 & 2.9\,$\pm$\,0.4 & 2.8\,$\pm$\,0.4 \\
 &  & Rapid & 16.9\,$\pm$\,1.7 & 10.1\,$\pm$\,1.6 & 1.6\,$\pm$\,1.2 \\
\cline{2-6}
 & \multirow[t]{3}{*}{disk} & Delayed & 134\,$\pm$\,11.1 & 67\,$\pm$\,10.5 & 33.4\,$\pm$\,10.4 \\
 &  & Rapid & 140.9\,$\pm$\,14.4 & 100.9\,$\pm$\,10.34 & 13.2\,$\pm$\,5.8 \\
\cline{1-6} \cline{2-6}
\multirow[t]{9}{*}{TESS} & \multirow[t]{3}{*}{bulge} & Delayed & 0.2\,$\pm$\,0.01 & 0.1\,$\pm$\,0.03 & 0.09\,$\pm$\,0.01 \\
 &  & Rapid & 0.3\,$\pm$\,0.05 & 0.1\,$\pm$\,0.1 & 0.06\,$\pm$\,0.09 \\
\cline{2-6}
 & \multirow[t]{3}{*}{disk} & Delayed & 24.0\,$\pm$\,0.1 & 8.6\,$\pm$\,1.1 & 1.2\,$\pm$\,0.7 \\
 &  & Rapid & 39.3\,$\pm$\,3.7 & 11.8\,$\pm$\,0.3 & 2.5\,$\pm$\,0.2 \\
\cline{1-6} \cline{2-6}
\multirow[t]{9}{*}{ZTF} & \multirow[t]{3}{*}{bulge} & Delayed & 2.1\,$\pm$\,0.1 & 1.8\,$\pm$\,0.2 & 1.0\,$\pm$\,0.3 \\
 &  & Rapid & 1.1\,$\pm$\,0.1 & 0.6\,$\pm$\,0.1 & 0.4\,$\pm$\,0.2 \\
\cline{2-6}
 & \multirow[t]{3}{*}{disk} & Delayed & 84.2\,$\pm$\,7.2 & 67.7\,$\pm$\,6.9 & 8.5\,$\pm$\,6.9 \\
 &  & Rapid & 62.1\,$\pm$\,8.1 & 47.2\,$\pm$\,6.1 & 9.1\,$\pm$\,5.42 \\
\cline{1-6} \cline{2-6}
\bottomrule
\end{tabular}
\caption{Detection rates of SL for surveys and Galactic components. Filter-only predictions ($n_\text{sl, filter only}$) are compared with SNR-constrained predictions requiring signal-to-noise ratios above 2$\sigma$ ($n_\mathrm{sl,2\sigma}$) or 3$\sigma$ ($n_\mathrm{sl,3\sigma}$). Uncertainties represent standard deviations from 12-fold bootstrapping.}
\label{table:raw_vs_sigma}
\end{table*}

Table~\ref{table:raw_vs_sigma} compares predicted stellar lensing event rates under different observational constraints: filter-only selections, where the flux is calculated within the instrumental band and SL is detected at $>$ zero sigma ($n_\text{sl, filter only}$), versus more conservative SNR thresholds ($n_\mathrm{sl,2\sigma}$ and $n_\mathrm{sl,3\sigma}$).

As expected, the results demonstrate substantial reductions in detectable events (by $\sim$two orders of magnitude) with respect to mere detections (Table~\ref{tab:results}). For LSST, the transition from filter-only to 3$\sigma$-limited detections reduces event rates by approximately 30–40$\%$ across all models and Galactic components. TESS exhibits the most severe reduction, with 3$\sigma$-limited predictions showing order-of-magnitude decreases compared to filter-only predictions, particularly for bulge observations. ZTF shows intermediate behaviour, with 3$\sigma$-limited predictions typically reduced by 50–60$\%$ relative to filter-only predictions.

Comparing the two population synthesis codes, \cosmic generally predicts higher event rates than \startrack, particularly for disk observations. For LSST disk observations, \cosmic predictions exceed \startrack by factors of 3 to 4. However, both codes show similar trends in the relative impact of SNR constraints, with the most stringent 3$\sigma$ threshold reducing detection rates substantially across all instruments as expected.

The impact of SNR constraints is particularly pronounced for mass gap systems, where $3\sigma$-limited predictions show dramatic reductions compared to unconstrained scenarios. For the f05/Delayed model, all three surveys converge to similar expectations of approximately $0.1\pm0.1$ detections. For the f40/Rapid model, the rates are $3.5\times10^{-3}$, $7.1\times10^{-3}$, and $2.8\times10^{-5}$ expected detections for LSST, ZTF, and TESS, respectively, highlighting the extreme difficulty of observing mass gap stellar lensing events with current survey capabilities and stringent detection thresholds.

\begin{table*}
\footnotesize
\begin{tabular}{lllrrrrrrrrrrrrrr}
\toprule
 &  &  & \multicolumn{2}{c}{$\langle \taueff\rangle$} & \multicolumn{2}{c}{$\langle \ml\rangle$} & \multicolumn{2}{c}{$\langle \ms\rangle$} & \multicolumn{2}{c}{$\langle \porb\rangle$} & \multicolumn{2}{c}{$\langle \ntot\rangle$} & \multicolumn{2}{c}{$\langle d\rangle$} & \multicolumn{2}{c}{$\langle \msource\rangle$} \\
 &  & $>3\sigma$ & No & Yes & No & Yes & No & Yes & No & Yes & No & Yes & No & Yes & No & Yes \\
instr & mwc & model &  &  &  &  &  &  &  &  &  &  &  &  &  &  \\
\midrule
\multirow[t]{6}{*}{LSST} & \multirow[t]{3}{*}{bulge} & f05 & 0.35 & 0.05 & 1.32 & 15.45 & 1.73 & 0.60 & 12.38 & 18.19 & 21.40 & 2.30 & 9.44 & 8.92 & 16.29 & 22.21 \\
 &  & f10 & 0.43 & 0.06 & 1.26 & 15.64 & 1.98 & 0.66 & 14.62 & 22.07 & 24.50 & 2.00 & 9.42 & 9.10 & 15.88 & 22.03 \\
 &  & f40 & 0.36 & 0.05 & 1.28 & 15.67 & 1.88 & 0.62 & 12.20 & 20.54 & 23.40 & 2.30 & 9.37 & 8.89 & 16.09 & 22.10 \\
\cline{2-17}
 & \multirow[t]{3}{*}{disk} & f05 & 0.33 & 0.05 & 1.32 & 14.20 & 2.21 & 0.64 & 11.02 & 17.04 & 22.30 & 2.70 & 9.43 & 8.93 & 15.59 & 21.80 \\
 &  & f10 & 0.34 & 0.05 & 1.24 & 15.11 & 2.15 & 0.70 & 11.05 & 14.26 & 22.35 & 2.90 & 9.48 & 9.17 & 15.66 & 21.66 \\
 &  & f40 & 0.34 & 0.06 & 1.28 & 14.84 & 2.20 & 0.70 & 10.56 & 17.41 & 22.40 & 2.60 & 9.37 & 9.19 & 15.54 & 21.52 \\
\cline{1-17} \cline{2-17}
\multirow[t]{6}{*}{TESS} & \multirow[t]{3}{*}{bulge} & f05 & 0.28 & 0.15 & 1.31 & 1.44 & 3.43 & 2.89 & 6.72 & 2.63 & 913.45 & 1124.23 & 4.81 & 0.07 & 12.88 & 5.13 \\
 &  & f10 & 0.35 & 0.04 & 1.19 & 1.28 & 4.08 & 3.01 & 6.32 & 4.91 & 1118.32 & 225.28 & 5.56 & 0.29 & 12.91 & 11.48 \\
 &  & f40 & 0.32 & 0.05 & 1.28 & 1.28 & 3.98 & 1.00 & 5.77 & 1.19 & 1170.63 & 431.58 & 5.50 & 0.41 & 13.00 & 12.37 \\
\cline{2-17}
 & \multirow[t]{3}{*}{disk} & f05 & 0.27 & 0.17 & 1.32 & 2.77 & 4.01 & 2.53 & 4.54 & 6.46 & 1183.92 & 602.48 & 5.48 & 0.66 & 13.03 & 9.87 \\
 &  & f10 & 0.27 & 0.15 & 1.19 & 10.85 & 4.05 & 2.82 & 4.64 & 5.77 & 1105.67 & 350.59 & 5.71 & 0.77 & 13.11 & 10.51 \\
 &  & f40 & 0.26 & 0.18 & 1.28 & 9.78 & 4.11 & 2.97 & 4.60 & 3.13 & 1162.16 & 494.34 & 5.81 & 0.87 & 13.09 & 10.12 \\
\cline{1-17} \cline{2-17}
\multirow[t]{6}{*}{ZTF} & \multirow[t]{3}{*}{bulge} & f05 & 0.18 & 0.05 & 1.45 & 15.60 & 1.68 & 0.63 & 2.45 & 31.26 & 103.90 & 3.40 & 9.31 & 7.21 & 16.93 & 21.09 \\
 &  & f10 & 0.21 & 0.07 & 1.26 & 16.25 & 1.88 & 0.66 & 4.14 & 38.53 & 96.96 & 4.10 & 9.24 & 7.09 & 16.36 & 21.32 \\
 &  & f40 & 0.19 & 0.06 & 1.28 & 15.73 & 1.80 & 0.64 & 3.03 & 34.21 & 111.00 & 3.80 & 9.23 & 7.05 & 16.63 & 21.36 \\
\cline{2-17}
 & \multirow[t]{3}{*}{disk} & f05 & 0.18 & 0.04 & 1.37 & 14.64 & 1.91 & 0.63 & 2.43 & 29.41 & 111.60 & 3.90 & 9.38 & 7.66 & 16.31 & 20.90 \\
 &  & f10 & 0.19 & 0.06 & 1.25 & 15.60 & 1.97 & 0.66 & 3.06 & 26.89 & 105.80 & 4.70 & 9.37 & 7.52 & 16.24 & 21.27 \\
 &  & f40 & 0.18 & 0.06 & 1.28 & 15.16 & 1.92 & 0.66 & 2.82 & 31.26 & 102.70 & 4.00 & 9.40 & 7.46 & 16.30 & 21.02 \\
\cline{1-17} \cline{2-17}
\bottomrule
\end{tabular}
\caption{Median system properties for SL events above and below the $3\sigma$ detection threshold. Systems are classified as detectable (Yes) or undetectable (No) based on SNR > $3\sigma$ for each survey-Galactic component combination. Physical parameters include: Einstein crossing time ($\langle \taueff\rangle$ [day]), lens mass ($\langle \ml\rangle\;[\msun]$]), source mass ($\langle \ms\rangle\; [\msun]$ ), orbital period ($\langle \porb\rangle$ [days]), total number of lensing data points ($\langle \ntot\rangle$), distance ($\langle d\rangle$ [kpc]), and source apparent magnitude ($\langle \msource\rangle$ [mag]). Results demonstrate systematic differences in the properties of detectable versus undetectable SL systems.}
\label{table:raw_vs_sigma_params}
\end{table*}

Table \ref{table:raw_vs_sigma_params} presents a comprehensive comparison of median system properties between events detectable above the 3$\sigma$ threshold (Yes) and those falling below this detection limit (No). The analysis reveals distinct systematic differences in the physical and observational characteristics of detectable versus undetectable systems.

Detectable systems consistently exhibit shorter effective Einstein  crossing times across all instruments and Galactic components. For LSST and ZTF, detectable systems show $\langle \taueff\rangle \sim 0.05-0.06$ [day] compared to $\langle \taueff\rangle \sim 0.18-0.43$ [day] for undetectable systems. TESS shows a less pronounced but similar trend, with detectable systems having $\langle \taueff\rangle \sim 0.04-0.18$ versus $\langle \taueff\rangle \sim 0.26-0.35$ for undetectable events.

A striking contrast emerges in lens masses: detectable systems are associated with significantly more massive BH lenses ($\langle \ml\rangle \sim 14-16 M_{\odot}$) compared to undetectable systems (NS with $\langle \ml\rangle \sim 1.2-1.5 M_{\odot}$). Conversely, detectable systems tend to have lower source masses, with $\langle \ms\rangle \sim 0.6-0.7 M_{\odot}$ versus $\langle \ms\rangle \sim 1.7-4.1 M_{\odot}$ for undetectable systems, although both are typically MS stars.

Detectable systems exhibit significantly longer orbital periods, particularly evident in ZTF observations where $\langle \porb\rangle \sim 26-39$ days for detectable systems compared to $\langle \porb\rangle \sim 2.4-4.1$ days for undetectable systems. LSST shows a similar but less extreme trend. The distance distributions show that detectable systems are generally located at smaller distances ($\langle d\rangle \sim 7-9$ kpc) compared to undetectable systems ($\langle d\rangle \sim 9.2-9.8$ kpc).

Detectable systems are associated with dimmer apparent source magnitudes ($\langle \msource\rangle \sim 20-22$ mag) compared to undetectable systems ($\langle \msource\rangle \sim 12-17$ mag). The situation is opposite for TESS where the detectable systems are typically brighter than the undetectable ones.

The total number of observational points shows instrument-specific patterns. For LSST and ZTF, detectable systems receive fewer total observations ($\langle \ntot\rangle \sim 2-5$) compared to undetectable systems ($\langle \ntot\rangle \sim 22$--$24$ or $90$--$110$ for LSST and ZTF, respectively). TESS exhibits the opposite trend with much higher observation totals for both categories.

These results highlight the complex interplay between system properties and detection thresholds in lensing surveys. The preference for massive lenses, shorter crossing times, and specific orbital configurations in detectable systems provides important constraints for optimizing survey strategies and interpreting observational results. The instrument-specific detection biases revealed in this analysis — particularly the systematic differences in observable system properties between detectable and undetectable events — must be carefully considered when comparing theoretical predictions with observational constraints from different survey programs. 

However, several factors could potentially increase these conservative detection rates significantly. Extended survey durations beyond the baseline observing programs would accumulate more photometric data points and improve signal-to-noise ratios for marginal events. More sophisticated Milky Way models, as demonstrated in W21, predict higher event rates through more accurate treatments of stellar populations and compact object distributions. Advances in noise reduction techniques and data processing algorithms could lower effective detection thresholds, while relaxing magnitude limits to include fainter sources would expand the observable target population. Furthermore, promising stellar lensing candidates identified through photometric surveys could be confirmed through follow-up radial velocity observations, converting marginal photometric detections into secure confirmations. These considerations suggest that our predictions represent conservative lower bounds.

\section{Discussion}

\subsection{Comparison with \citet{Wiktorowicz_2021}}

Our study extends the work of W21 by specifically investigating how different SN engines influence the population of compact objects in binary systems detectable through SL. While we employed a simplified Galactic model consisting of only the thin disk and bulge components, this simplification is justified, as W21 demonstrated that other Galactic components contribute minimally to the overall mass and do not significantly affect SL detection rates. Similarly, our adoption of the SFH and chemical evolution model from \citet{Olejak2020} streamlines calculations without compromising the validity of our conclusions.

Both studies reach similar conclusions regarding the relative detection capabilities of different surveys. ZTF and LSST consistently emerge as more promising platforms for SL detections compared to TESS (cf. table 2 in W21), primarily due to their extended survey durations and superior sensitivities.

The detection rates predicted by our study, particularly for the disk population under the f05 model, fall within the lower range of the broader predictions presented in W21 for various IMFs. This quantitative difference likely stems from our simplified Galactic model and specific evolutionary assumptions, compared to the more detailed Milky Way model employed in W21.

Our analysis reveals a strong preference for SL systems with low centre-of-mass velocities ($\vcm$), consistent with the physical understanding that binaries experiencing significant natal kicks have a lower probability of remaining bound and subsequently detectable. While W21 implicitly incorporated natal kick effects on binary survival, our study explicitly examines $\vcm$ distributions across different SN models, providing new insights into this relationship.

In summary, while our results align with W21 regarding overall survey detection trends, our study contributes novel insights into aspects not previously explored in depth. Specifically, we illuminate the critical influence of different SN formation mechanisms on SL detectability and characterize the properties of SL systems within the mass gap. These findings, enabled by our focused methodology, complement the broader scope of W21 and advance our understanding of SL binary populations.

\subsection{Comparison between \startrack and \cosmic results}\label{sec:codes_comparison}

Our analysis employs two independent binary population synthesis codes to assess the robustness of our SL predictions and quantify inter-code systematic uncertainties. The specific implementations of the same physical processes may differ and be subject to different optimization. While both codes share similar evolutionary prescriptions, several key differences in their implementations provide valuable insights into model dependencies and systematic uncertainties in our results. 

The primary differences between \startrack and \cosmic lie in their treatment of specific evolutionary phases and physical prescriptions. \cosmic adopts a slightly extended IMF, incorporating an additional low-mass regime ($\alpha_0 = -0.3$ for $\mathrm{M} \in [0.01, 0.08]\msun$) compared to the standard broken power-law used in \startrack. The secondary mass sampling also differs, with \cosmic employing a uniform mass ratio distribution constrained by pre-main sequence lifetime considerations, while \startrack uses its own prescription for binary mass ratios.

The SN mechanisms represent the most significant difference in our comparison. \startrack employs three convective mixing models \citep[f05, f10, f40 corresponding to $\fmix = 0.5, 1.0, 4.0$;][]{Fryer2022}, while \cosmic uses the "Delayed" and "Rapid" explosion models \citep{Fryer1204} that roughly correspond to the f05 and f40 extremes, respectively.

The Rapid convective growth models (f40/Rapid) consistently create a prominent mass gap between 2–5$\msun$ in both codes, while the delayed models (f05/Delayed) efficiently populate this region with massive NSs and low-mass BHs.

Minor differences in the detailed shape of the mass distributions of lenses and sources likely reflect the subtle differences in evolutionary prescriptions and numerical implementations between the codes. However, these variations are well within the expected range of systematic uncertainties inherent in population synthesis modelling.

Table \ref{tab:results} reveals that \cosmic generally predicts fewer detectable SL events than \startrack, particularly for disk populations. The most significant discrepancy occurs in the ZTF disk predictions, where \cosmic yields approximately $60\%$ fewer events than the \startrack f05 model ($2,700$ vs $6,300$ events). However, \cosmic results align more closely with the \startrack f40 model, suggesting that the delayed vs rapid SN mechanisms represent the primary source of inter-code variation. For bulge populations, the agreement between codes is much better.

The ranking of surveys by detection potential remains consistent across both codes: ZTF shows the highest numbers, followed by LSST, then TESS. This reflects the same prescriptions used in post-processing for analysis of both code results.

The consistent qualitative trends across both codes -- including the survey ranking, disk/bulge ratios, and parameter distribution shapes -- provide confidence in our main conclusions. The quantitative differences, while non-negligible, do not alter the fundamental predictions that ZTF can detect large numbers of SL events, with disk populations dominating the observable sample.

\section{Conclusions}

Using the \startrack\ and \cosmic\ population synthesis codes, we have investigated how different supernova engines affect the observable population of SL binary systems in the Milky Way. Our analysis reveals several key findings that advance our understanding of compact object populations and their detectability through SL.

Our models predict substantial numbers of detectable SL events across major astronomical surveys. For the Galactic bulge, we expect 21-650 events, while the Galactic disk should yield 370-6,300 events depending on the survey and SN model. ZTF consistently shows the highest detection potential, followed by LSST, then TESS. This ranking remains robust across all SN models and reflects fundamental differences in assumed survey duration, cadence, and sky coverage.

The predicted SL event rates show relatively modest variations (factors of $\lesssim2$) across different SN models for most configurations. However, we identify a notable exception in the Galactic disk population, where the $\fmix=0.5$ (Delayed) model predicts substantially higher event counts -- up to a factor of four more than the $\fmix=4.0$ (Rapid) model for certain surveys. This difference between disk and bulge populations suggests that SL observations could serve as a diagnostic tool for constraining SN mechanisms.

A key finding of our study is that the $\fmix=0.5$ model predicts a significantly higher number of SL systems with lens masses in the controversial $2$--$5\msun$ mass gap region. ZTF shows particularly strong discriminating power, with the $\fmix=0.5$ model predicting approximately $10$ times more mass gap detections than the $\fmix=1.0$ model in the Galactic disk. This suggests that SL observations could provide crucial constraints on the existence and properties of compact objects in this mass range, potentially helping to resolve current debates about the presence and occupation of the mass gap.

Our analysis reveals three distinct populations of SL systems based on lens mass and Einstein crossing time: low-mass systems (predominantly NSs with $\mlens \leq 2 \msun$), and two BH populations distinguished by their orbital periods. The \gns\ population exhibits the highest maximum magnifications and lowest centre-of-mass velocities (due to formation through electron-capture SN), while both BH populations show higher velocities. These natural groupings provide a framework for system classification and targeted follow-up strategies.

SL systems exhibit a strong preference for low centre-of-mass velocities ($\vcm \lesssim 20$ km/s) across all models and instruments. This characteristic results from fundamental selection effects: binaries that experience large natal kicks during SN explosions are less likely to remain gravitationally bound. The observed velocity distributions therefore reflect the physical processes governing compact object formation and binary survival, with massive BHs and those formed in electron-capture SN being preferentially retained.

While the majority of predicted SL events will have limited observational coverage, our models identify a valuable subset of targets with multiple observations and recurrent events ($\ntot > 50$). However, only a handful of them will be detectable above $3\sigma$ detection limit. The enhanced discriminating power between SN models for well-observed systems underscores the scientific value of long-duration, high-cadence surveys.

The comparison between \startrack\ and \cosmic\ codes reveals generally consistent results, with \cosmic's Delayed and Rapid models producing outcomes comparable to \startrack's $\fmix=0.5$ and $\fmix=4.0$ models, respectively. However, \cosmic\ typically predicts somewhat more events, particularly for bulge populations, highlighting the importance of systematic uncertainties in population synthesis modelling. The consistent qualitative trends across both codes — including survey rankings, disk-to-bulge ratios, and parameter distributions — lend confidence to our main conclusions.

When applying realistic observational constraints including photometric precision and signal-to-noise requirements, predicted detection rates decrease by approximately two orders of magnitude compared to raw estimates. LSST maintains the best performance under conservative assumptions, with $3\sigma$-limited predictions showing only $30$--$40\%$ reductions from filter-only selections. These estimates provide conservative expectations for upcoming surveys and highlight the importance of definitively identifying SL events above instrumental noise. We note that extended survey durations, more sophisticated Milky Way models, advances in noise reduction techniques and data processing algorithms, and relaxing magnitude limits to include fainter sources, may significantly increase the observable target population.

Our results demonstrate that SL observations represent a promising avenue for probing both SN physics and the Galactic population of compact objects. The predicted detection rates, particularly for ZTF and LSST, are sufficiently high to enable statistical studies of compact object properties. Most importantly, the enhanced sensitivity to mass gap objects in f05 or Delayed SN models (up to $\sim45$ expected detections for ZTF) suggests that SL surveys could provide the first definitive observational constraints on this controversial population, thereby informing our understanding of stellar collapse and compact object formation mechanisms.

\section*{Acknowledgments}
 
GW was supported by the Polish National Science Center (NCN) through the grant 2021/41/B/ST9/01191. MM acknowledges support from STFC Consolidated grant (ST/V001000/1). AI acknowledges support from the Royal Society. For the purpose of Open Access, the authors have applied for a CC-BY public copyright license to any Author Accepted Manuscript (AAM) version arising from this submission. 

\section*{Data Availability}

Data available on request. 


\bibliographystyle{mnras}
\bibliography{ms}





\bsp	
\label{lastpage}
\end{document}